\documentclass[a4paper,fleqn,usenatbib]{mnras}

\usepackage{newtxtext,newtxmath}
\usepackage[T1]{fontenc}

\usepackage[english]{babel}
\usepackage{graphicx}
\usepackage{fleqn}
\usepackage{amssymb}
\usepackage{amsmath}
\usepackage{booktabs}
\usepackage{hyperref}
\usepackage{comment}

\usepackage{color, colortbl}

\newcommand*{\mycdot}{\kern-.2em\cdot\kern-.2em}
\renewcommand{\S}{Section}
\newcommand{\F}{Fig.}
\newcommand{\codename}{SecularMultiple}

\newcommand{\ve}[1]{\boldsymbol{#1}}
\newcommand{\unit}[1]{\hat{\boldsymbol{#1}}}

\newcommand{\msun}{\mathrm{M}_\odot}

\newcommand{\au}{\,\textsc{au}}
\newcommand{\renc}{R_\mathrm{enc}}
\newcommand{\srel}{\sigma_\mathrm{rel}}
\newcommand{\mper}{M_\mathrm{per}}

\pubyear{2018}

\allowdisplaybreaks

\title[SNe Ia in quadruples]{On the rates of type Ia supernovae originating from white dwarf collisions in quadruple star systems}

\author[Hamers]{Adrian S. Hamers$^{1}$\thanks{E-mail: hamers@ias.edu} \\
$^{1}$Institute for Advanced Study, School of Natural Sciences, Einstein Drive, Princeton, NJ 08540, USA}

\date{Accepted 2018 April 13. Received 2018 April 05; in original form 2018 March 07}

\begin{document}

\label{firstpage}
\pagerange{\pageref{firstpage}--\pageref{lastpage}}
\maketitle

\begin{abstract} 
We consider the evolution of stellar hierarchical quadruple systems in the 2+2 (two binaries orbiting each other's barycentre) and 3+1 (triple orbited by a fourth star) configurations. In our simulations, we take into account the effects of secular dynamical evolution, stellar evolution, tidal evolution and encounters with passing stars. We focus on type Ia supernovae (SNe Ia) driven by collisions of carbon-oxygen (CO) white dwarfs (WDs). Such collisions can arise from several channels: (1) collisions due to extremely high eccentricities induced by secular evolution, (2) collisions following a dynamical instability of the system, and (3) collisions driven by semisecular evolution. The systems considered here have initially wide inner orbits, with initial semilatus recti larger than $12\,\au$, implying no interaction if the orbits were isolated. However, taking into account dynamical evolution, we find that $\approx 0.4$ ($\approx 0.6$) of 2+2 (3+1) systems interact. In particular, Roche Lobe overflow can be triggered possibly in highly eccentric orbits, dynamical instability can ensue due to mass-loss-driven orbital expansion or secular evolution, or a semisecular regime can be entered. We compute the delay-time distributions (DTDs) of collision-induced SNe Ia, and find that they are flatter compared to the observed DTD. Moreover, our combined SNe Ia rates are $(3.7\pm0.7) \times 10^{-6} \,\mathrm{M}_\odot^{-1}$ and $(1.3\pm0.2) \times 10^{-6} \,\mathrm{M}_\odot^{-1}$ for 2+2 and 3+1 systems, respectively, three orders of magnitude lower compared to the observed rate, of order $10^{-3} \, \mathrm{M}_\odot^{-1}$. The low rates can be ascribed to interactions before the stars evolve to CO WDs. However, our results are lower limits given that we considered a subset of quadruple systems.
\end{abstract}


\section{Introduction}
\label{sect:introduction}
Type Ia supernovae (SNe Ia) are of key importance in astrophysics. Their relatively uniform luminosity allows for distance determination on cosmological scales, which has provided evidence for the accelerated expansion of the universe \citep{1998AJ....116.1009R,1999ApJ...517..565P}. However, the origin of SNe Ia is not well established (see, e.g., \citealt{2012NewAR..56..122W,2014ARA&A..52..107M,2018arXiv180203125L} for reviews). It is believed that SNe Ia are associated with runaway thermonuclear explosions of degenerate carbon-oxygen (CO) white dwarfs (WDs). These explosions might arise if a WD accretes matter from a companion in a binary and exceeds the Chandrasekhar mass (the `single degenerate', SD, channel, \citealt{1973ApJ...186.1007W,1984ApJ...286..644N}), or if two WDs merge in a tight system due to gravitational wave emission after common-envelope evolution (the `double degenerate', DD, channel, \citealt{1984ApJS...54..335I,1984ApJ...277..355W}). Both these channels face a number of problems, including the apparent absence of close non-degenerate companions in the case of the SD channel, and the low predicted rates in the case of the DD channel. 

Another channel for SNe Ia has been given attention which involves collisions of WDs (i.e., `violent' mergers). Colliding WDs are potentially efficient SNe Ia sources (e.g., \citealt{2009MNRAS.399L.156R,2009ApJ...705L.128R,2010ApJ...724..111R,2012ApJ...747L..10P,2015ApJ...807..105S}), but the collision rates in dense stellar systems such as globular clusters are expected to be too low compared to observed SNe Ia rates (e.g., \citealt{2009ApJ...705L.128R}). Instead, it has been suggested that WD collisions can occur in triple star systems \citep{2011ApJ...741...82T}, and that their rates are potentially high \citep{2012arXiv1211.4584K}. In this scenario, the inner binary, consisting of two WDs and orbited by a third object (star or compact object), is driven to very high eccentricity by the torque of the tertiary object through Lidov-Kozai (LK) oscillations (\citealt{1962P&SS....9..719L,1962AJ.....67..591K}; see \citealt{2016ARA&A..54..441N} for a review). However, the inner system is typically wide due to the stellar evolution that preceded, and a very high eccentricity is required for the WDs to collide. This in turn requires a high mutual inclination, close to $90^\circ$, but the latter implies that a large fraction of potential progenitor systems merges during earlier stages of stellar evolution (in particular, during the main sequence, MS, or giant stages), implying that the rates are low \citep{2013MNRAS.430.2262H,2017arXiv170900422T}.

Recently, \citet{2017arXiv170908682F} considered a variation on the WD collision scenario in triples: WD collisions in stellar quadruple systems composed of two binaries orbiting each other's barycentre. Although quadruple systems are less common than triple systems (for Solar-type stars, triples are about 10 times more common than quadruples, \citealt{2014AJ....147...86T,2014AJ....147...87T}), the efficiency of attaining high eccentricities in these quadruple systems is higher compared to triples (e.g., \citealt{2013MNRAS.435..943P}), and \citet{2017arXiv170908682F} argued that the higher efficiency implies that the quadruple WD collision rate, and hence the SNe Ia rate, is interestingly high. 

However, the same arguments of \citet{2013MNRAS.430.2262H} and \citet{2017arXiv170900422T} might apply to hierarchical quadruple systems: in order for the WDs to collide, a highly fine-tuned system is required and many progenitor systems merge before a double WD binary is formed, implying low rates. The problem is complicated, however, by the fact that the dynamics in quadruple systems are more complex compared to triples: the long-term evolution can be chaotic even to lowest order \citep{2017MNRAS.470.1657H}, meaning that the time-scale for reaching high eccentricities can be long, i.e., longer than the MS lifetimes. 

In this paper we address this issue and study, using population synthesis calculations, the long-term evolution of quadruple systems with stellar masses in the range relevant for CO WDs (between 1 and 6.5 $\msun$), with the goal of estimating the rate of WD collisions, and hence of potential SNe Ia. We take into account the secular (i.e., orbit-averaged) dynamical evolution, stellar evolution (including the effects of mass loss on the orbits), tidal evolution (taking into account the changing structure of the stars), and flybys from passing stars in the field. Also, we take into account the semisecular regime in which the averaged equations of motion break down \citep{2012arXiv1211.4584K}. We consider quadruple systems in both the `2+2' (two binaries orbiting each other's barycentre) and `3+1' (a triple orbited by a fourth star) configurations. To our knowledge, this is the first time that these processes are taken into account simultaneously in the long-term evolution of stellar quadruple systems.

In \S\,\ref{sect:meth}, we describe the numerical algorithm used for our simulations. The initial conditions and assumptions for the population synthesis are discussed in \S\,\ref{sect:IC}. In \S\,\ref{sect:examples}, we illustrate several evolutionary pathways found in the simulations. We present our results in \S\,\ref{sect:results}, discuss them in \S\,\ref{sect:discussion}, and conclude in \S\,\ref{sect:conclusions}.

\section{Numerical algorithm}
\label{sect:meth}

\begin{figure}
\center
\includegraphics[scale = 0.65, trim = 0mm 0mm 0mm 0mm]{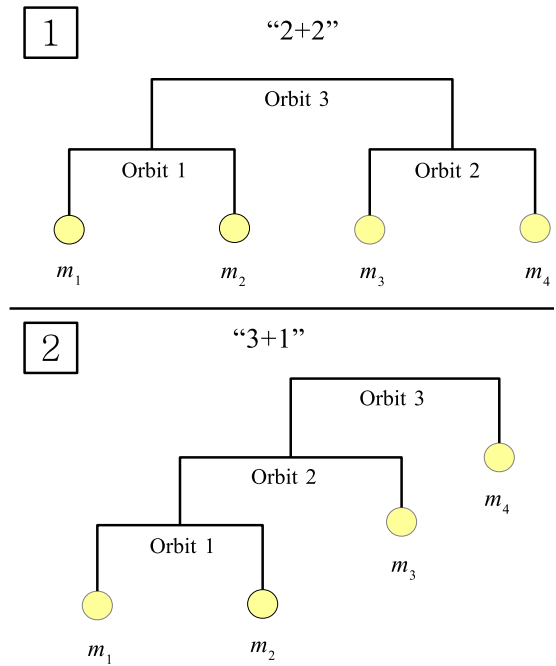}
\caption {Schematic depiction of the types of systems considered in this paper in a mobile diagram \citep{1968QJRAS...9..388E}. Top: the 2+2 configuration; bottom: the 3+1 configuration. }
\label{fig:configurations}
\end{figure}

\begin{table*}
\begin{tabular}{lp{6.0cm}p{8.0cm}}
\toprule
Symbol & Description & Initial value(s) and/or distribution in population synthesis \\
\midrule
{\it Stellar quadruple system} \\
$m_1$					& Mass of the primary star in orbit 1.																& $1-6.5\,\msun$ with a Kroupa initial mass function (\citealt{1993MNRAS.262..545K}, i.e., $\mathrm{d} N/\mathrm{d}m_1 \propto m_1^{-2.7}$ in this mass range). \\
$m_2$					& Mass of the secondary star in orbit 1. 															& $m_1 q_1$, where $q_1$ has a flat distribution between 0.01 and 1, and with $m_2>1\,\msun$. \\
$m_3$ (2+2)				& Mass of the primary star in orbit 2 (2+2). 														& $q_3 (m_1+m_2) /(1+q_2)$, where $q_2$ and $q_3$ have flat distributions between 0.01 and 1, and with $m_3>0.1\,\msun$. \\
$m_3$ (3+1)				& Mass of the star orbiting the inner binary (3+1).													& $q_2 (m_1+m_2)$, where $q_2$ has a flat distribution between 0.01 and 1, and with $m_3>0.1\,\msun$. \\
$m_4$ (2+2)				& Mass of the secondary star in orbit 2 (2+2).														& $q_2 m_3$, where $q_2$ was sampled for $m_3$, and with $m_4>0.1\,\msun$. \\
$m_4$ (3+1)				& Mass of the outermost star in orbit 3 (3+1).														& $q_3 m_3$, where $q_3$ has a flat distribution between 0.01 and 1, and with $m_4>0.1\,\msun$. \\
$Z_i$					& Metallicity of star $i$.																		& $0.02$ \\
$R_i$					& Radius of star $i$.																			& From stellar evolution code. \\
$P_{\mathrm{s},i}$			& Spin period of star $i$. 																		& $10\,\mathrm{d}$ \\
$\theta_{\mathrm{s},i}$		& Obliquity (spin-orbit angle) of star $i$. 															& $0^\circ$ \\
$t_{\mathrm{V},i}$			& Viscous time-scale of star $i$. 																& Computed from the stellar properties using the prescription of \citet{2002MNRAS.329..897H}. \\
$k_{\mathrm{AM},i}$			& Apsidal motion constant of star $i$.															& 0.014 \\
$r_{\mathrm{g},i}$			& Gyration radius of star $i$.																	& 0.08 \\
$P_{\mathrm{orb},i}$			& Orbital period of orbit $i$. 																	& Gaussian distribution in $\mathrm{log}_{10}(P_{\mathrm{orb},i}/\mathrm{d})$ with mean 5.03 and standard deviation 2.28 \citep{2010ApJS..190....1R}, subject to dynamical stability constraints \citep{2001MNRAS.321..398M}, and $a_i(1-e_i^2) > 12 \, \au$ for orbits 1 and 2 (2+2), or orbit 1 (3+1). \\
$a_i$					& Semimajor axis of orbit $i$. 																	& Computed from $P_{\mathrm{orb},i}$ and the $m_i$ using Kepler's law. \\
$e_i$					& Eccentricity of orbit $i$. 																		& Rayleigh distribution between 0.01 and 0.9 with an rms width of 0.33 \citep{2010ApJS..190....1R}. \\
$i_i$						& Inclination of orbit $i$. 																		& $0-180^\circ$ (flat distribution in $\cos i_i $) \\
$i_{ij}$					& Inclination of orbit $i$ relative to orbit $j$.														& $0-180^\circ$ (flat distribution in $\cos i_{ij} $) \\
$\omega_i$				& Argument of periapsis of orbit $i$. 																& $0-360^\circ$ (flat distribution in $\omega_i$) \\
$\Omega_i$				& Longitude of the ascending node of orbit $i$. 														& $0-360^\circ$ (flat distribution in $\Omega_i$) \\
\midrule
{\it Flybys} \\
$M_\mathrm{per}$			& Mass of the perturbers. 																		& $0.1-80\,\msun$ with a Kroupa initial mass function \citep{1993MNRAS.262..545K}, corrected for gravitational focusing and a stellar age of 10 Gyr. \\
$n_\star$					& Stellar number density. 																		& $0.1 \, \mathrm{pc^{-3}}$ \citep{2000MNRAS.313..209H} \\
$\renc$					& Encounter sphere radius.																	& $0, 10^4\,\au$ \\
$\sigma_\star$				& One-dimensional stellar velocity dispersion.														& $30\,\mathrm{km\,s^{-1}}$ \\
\bottomrule
\end{tabular}
\caption{Description of important quantities and their initial value(s) and/or distributions assumed in the population synthesis. }
\label{table:IC}
\end{table*}

We model the long-term evolution of stellar hierarchical quadruple systems in both the `2+2' (two binaries orbiting each other's barycentre) and `3+1' (a triple orbited by a fourth star) configurations. A schematic depiction is shown in \F\,\ref{fig:configurations}. An overview of the notation used in this paper is given in the first two columns of Table~\ref{table:IC}. Our numerical algorithm is implemented within the \textsc{AMUSE} framework \citep{2013CoPhC.183..456P,2013A&A...557A..84P}, and below we describe the various ingredients in the code.

\subsection{Secular dynamical evolution}
\label{sect:meth:sec}
For the secular dynamics, we use \textsc{\codename} \citep{2016MNRAS.459.2827H}, which is a generalisation of a code developed earlier for 3+1 quadruple systems \citep{2015MNRAS.449.4221H}. The \textsc{\codename} code is based on an expansion of the Hamiltonian of the system in terms of ratios of separations of binaries on adjacent levels. The Hamiltonian is subsequently orbit averaged, and the orbit-averaged equations of motion are solved numerically. In the integrations, we include terms up to and including octupole order (third order in the separation ratios) for interactions involving three binaries, and up to and including dotriacontupole order (fifth order in the separation ratios) for pairwise interactions. 

Post-Newtonian (PN) corrections are included in each orbit to the 1 and 2.5PN orders (i.e., relativistic precession, and energy and angular-momentum loss due to gravitational wave radiation). Any `cross' terms, i.e., PN terms involving more than one orbit simultaneously \citep{2013ApJ...773..187N}, are neglected.

\subsection{Stellar evolution}
\label{sect:meth:stev}
The secular code is coupled within \textsc{AMUSE} with the stellar evolution code \textsc{SeBa} \citep{1996A&A...309..179P,2012A&A...546A..70T}, which is based on analytic fits to detailed stellar evolution calculations. The stellar evolution code is used to compute the masses and radii of the stars, which are assumed to start on the zero-age MS, as a function of time. We set the metallicity of all stars to $Z_i=0.02$. 

The stellar mass loss induces changes on the orbits. Throughout, we assume isotropic and adiabatic mass loss to compute the dynamical response of the orbits on mass loss, i.e., $a_i M_i$ and $e_i$ are constant \citep{1956AJ.....61...49H,1963ApJ...138..471H}, where $a_i$ and $e_i$ are the semimajor axis and eccentricity, respectively, of orbit $i$, and $M_i$ is the mass of all bodies contained within orbit $i$.

The effects of mass and radius changes due to stellar evolution on the stellar spins are taken into account assuming conservation of spin angular momentum, i.e., $m_i R_i^2 \Omega_i$ is constant due to these processes. In particular, this implies significant stellar spin down during the giant stages, and spin up at the formation of a WD. In some cases, conservation of spin angular momentum would imply a WD spin rate which is faster than the critical rotation rate, $\Omega_\mathrm{crit} = \sqrt{Gm_i/R_i^3}$. In that case, we set the WD spin rate to $\Omega_\mathrm{crit}$.

Furthermore, we assume that the obliquity $\theta_{\mathrm{s},i}$, i.e., the stellar-spin-orbit angle, is not affected by mass and radius changes due to stellar evolution (note that  $\theta_{\mathrm{s},i}$ can change due to orbital and/or tidal spin evolution).

\subsection{Tidal evolution}
\label{sect:meth:tides}
Tidal evolution is modelled with the equilibrium tide model \citep{1981A&A....99..126H,1998ApJ...499..853E}, using the equations of \citet{2009MNRAS.395.2268B}. We include the effects of tidal dissipation and spin-orbit coupling (precession due to tidal bulges and stellar rotation). The spin vectors of all stars are tracked, and the spins are not confined to be parallel with the orbit, although we initialise the spins to be parallel with the orbit (i.e., the initial obliquity $\theta_{\mathrm{s},i}=0^\circ$), with a spin period of $P_{\mathrm{s},i} = 10\,\mathrm{d}$. 

The tidal dissipation strength (i.e., the viscous time-scale or another equivalent tidal time-scale) is computed as a function of the stellar properties (in particular, the stellar type, mass, radius, convective envelope mass and radius, and stellar spin period) using the prescription of \citet{2002MNRAS.329..897H}, with a fixed apsidal motion constant of $k_{\mathrm{AM},i}=0.014$, and gyration radius of 0.08 (the latter values similar to \citealt{2007ApJ...669.1298F}). 

We remark that many uncertainties exist regarding the efficiency of tidal dissipation (see, e.g., \citealt{2014ARA&A..52..171O} for a review). In particular, we do not take into account dynamical tides, which could be important at high eccentricities and affect our results. Also, the prescription of \citet{2002MNRAS.329..897H} should be regarded to give a rough estimate of the efficiency of tidal dissipation in the equilibrium tide model, at best. However, the prescription does provide an estimate for all stages of stellar evolution, and, in this regard, is a better alternative than simply assuming a fixed viscous time-scale. In particular, the efficiency of tidal dissipation is much larger during the giant stages due to the presence of large convective envelopes, and this effect is taken into account in the prescription of \citet{2002MNRAS.329..897H}.

\subsection{Flybys}
\label{sect:meth:flybys}
We include the effects of passing stars on the system by considering impulsive encounters, i.e., encounters for which the relative motion is much faster than the orbital motion. In this approximation, the stars in the quadruple system can be considered to be fixed in space whereas the perturber imparts velocity kicks on each of the components. These kicks lead to changes of the orbits, affecting all orbital elements, in particular the semimajor axes and eccentricities. Due to the relatively low orbital speeds, the kicks are most important for the widest orbit in the system (i.e., orbit 3). The approximation is not strictly correct for tight orbits for which the orbital speed is potentially higher than the speed of the perturbing star. However, we restrict our population synthesis to relatively wide orbits (wider than $12\,\au$), for which the typical orbital speeds ($\lesssim 10\,\mathrm{km\,s^{-1}}$), are lower than the typical encounter speed ($\sim 30\,\mathrm{km\,s^{-1}}$), such that the majority of the encounters are impulsive.

We remark that the effects of flybys on the rates of WD mergers in triples have been discussed in S6.1 of \citet{2016MNRAS.456.4219A}.

\subsubsection{Effect on the orbits}
We use the routines to incorporate instantaneous orbital changes described in \citet{2018arXiv180205716H} appropriate for impulsive encounters. Specifically, assuming the perturber (mass $M_\mathrm{per}$) moves on a straight line with velocity $\ve{V}_\mathrm{per}$ with respect to the barycentre of the quadruple system, the velocity kick on body $i$ is given by
\begin{align}
\label{eq:imp}
\Delta \ve{V}_i &= 2 \frac{GM_\mathrm{per}}{V_\mathrm{per}} \frac{\unit{b}_i}{b_i},
\end{align}
where the impact parameter vector of body $i$,
\begin{align}
\ve{b}_i \equiv \ve{b}-\ve{R}_i - \unit{V}_\mathrm{per} \left [ \left (\ve{b}-\ve{R}_i \right ) \mycdot \unit{V}_\mathrm{per} \right ],
\end{align}
is defined in terms of the impact parameter vector $\ve{b}$. In practice, this gives a negligible effect on orbits 1 and 2, whereas orbit 3 can be significantly perturbed in terms of its orbital elements.

\subsubsection{Sampling the perturbers}
We use a methodology similar to \citet{2017AJ....154..272H} to sample passing stars in the simulations. In particular, we assume a locally-homogeneous stellar background with a number density $n_\star = 0.1 \, \mathrm{pc^{-3}}$ \citep{2000MNRAS.313..209H}, a one-dimensional velocity dispersion $\sigma_\star = 30\,\mathrm{km\,s^{-1}}$, and a Kroupa mass function \citep{1993MNRAS.262..545K} between 0.1 and 80 $\msun$, corrected for gravitational focusing and stellar evolution. The correction for stellar evolution is carried out by replacing the initial mass with the final mass after 10 Gyr of stellar evolution using the same stellar evolution code as described in \S\,\ref{sect:meth:stev}. 

We consider stars impinging on from a random direction $\unit{R}_\mathrm{enc}$ on the `encounter sphere' with radius $\renc$, and sample a velocity of the perturber relative to the barycentre of the quadruple system, $\ve{V}_\mathrm{per}$, assuming a Maxwellian distribution and taking into account an additional factor of $V_{\mathrm{per},z}$ (in the direction of $-\unit{R}_\mathrm{enc}$) to ensure the correct flux into the encounter sphere \citep{1972A&A....19..488H}. The impact parameter is then computed according to
\begin{align}
\ve{b} = \ve{R}_\mathrm{enc} - \left ( \unit{V}_\mathrm{per} \mycdot \ve{R}_\mathrm{enc} \right ) \unit{V}_\mathrm{per},
\end{align}
which implies velocity kicks to the bodies given by equation~(\ref{eq:imp}). 

The next perturber is sampled assuming that the probability for a time delay between encounters exceeding $\Delta t$ is $\mathrm{exp}(- \Gamma \Delta t)$, where the encounter rate $\Gamma$ is given by
\begin{align}
\label{eq:Gamma}
\nonumber \Gamma &= 2 \sqrt{2\pi} \renc^2 n_{\star} \srel \\
&\quad \times \int \mathrm{d} \mper f(\mper ) \left [1 + \frac{G(M_\mathrm{int}+\mper)}{\renc \srel^2} \right ],
\end{align}
with $f(\mper)\,\mathrm{d}\mper$ the fraction of perturbing stars with masses in the interval $\mathrm{d} \mper$, $M_\mathrm{int} = m_1+m_2+m_3+m_4$ the total mass of the quadruple system, and $\srel = \sqrt{2} \, \sigma_\star$ the relative velocity dispersion \citep{2008gady.book.....B}. 

The encounter radius is set to $\renc = 10^4\,\au$, such that most encounters with the widest orbit (orbit 3) are impulsive, whereas not too large as to be computationally too inhibitive. If a sampled encounter with respect to the widest orbit is not impulsive, i.e., if 
\begin{align}
\frac{\dot{f}_\mathrm{per}}{n_3} \approx \left [ \left ( 1 + \frac{M_\mathrm{per}}{M_\mathrm{int}} \right ) \left (\frac{a_3}{b} \right )^3 \left ( 2 + \frac{b V_\mathrm{per}^2}{G(M_\mathrm{per}+M_\mathrm{int})} \right ) \right ]^{1/2} \leq 1,
\end{align}
where $\dot{f}_\mathrm{per}$ is the angular speed of the perturber at periapsis (assuming a hyperbolic orbit) and $n_3$ is the mean motion of orbit 3, then we reject it. With this approach the secular encounters are neglected. Note, however, that, for wide orbits in the Solar neighbourhood, the effects of secular encounters are typically negligible compared to those of impulsive encounters (Hamers, unpublished).

\subsection{Stopping conditions}
\label{sect:meth:sc}
Clearly, our algorithm treats the evolution of the quadruple system in a simplified way and cannot model all facets of the evolution. We impose a number of stopping conditions to ensure that the simulation does not continue into a regime in which the underlying assumptions completely break down. In particular, we stop the simulations at the onset of Roche lobe overflow (RLOF), since in our case RLOF is typically triggered in eccentric orbits in which secular eccentricity driving is important (see \S\,\ref{sect:results:orb:e} below). The effects of RLOF in eccentric orbits with potentially similar mass transfer and secular time-scales, although interesting in their own right, are beyond the scope of this paper. Most importantly, however, a common consequence of RLOF is common-envelope (CE) evolution, in which case the orbit shrinks by orders of magnitude, and the hierarchy of the system increases significantly (i.e., the ratio of the semimajor axis of the innermost orbit to that of its parent decreases), implying that subsequent secular evolution is unimportant (see, e.g., \citealt{2013MNRAS.430.2262H} for an example of this phenomenon for triples). 

Furthermore, we stop the simulations when the secular approximation breaks down. The latter can occur when the angular-momentum time-scale becomes comparable or even shorter than the orbital periods, in which case averaging over the orbits is clearly no longer a good approximation. This regime, known as the quasi-secular or semisecular regime \citep{2014ApJ...781...45A}, is associated with typically higher eccentricities compared to the expectation from fully orbit-averaged secular theory. Another regime is that of dynamical instability, triggered if the system becomes more compact, and in which case stars may collide or are ejected from the system. 

Specifically, we impose the following stopping conditions. 
\begin{enumerate}
\item One of the stars fills its Roche lobe, potentially in an eccentric orbit. To determine the Roche lobe radius, we use the fits of \citet{2007ApJ...660.1624S}, in particular, equations~(47) through (52) evaluated at periapsis. The latter equations give the Roche lobe radius as a function of orbital phase, spin frequency, eccentricity and mass ratio, as a correction to the Roche lobe radius fits of \citet{1983ApJ...268..368E} evaluated at periapsis (i.e., $a$ in \citealt{1983ApJ...268..368E} is replaced by $a[1-e]$). We do not check for RLOF if the star has evolved to a WD. In that case, we check for physical collision at periapsis, i.e., if the sum of the radii of the objects in the orbit exceeds $a_i(1-e_i)$.
\item One of the orbit pairs becomes dynamically unstable. To evaluate stability, we use the criterion of \citet{2001MNRAS.321..398M}, which is assumed to be correct for quadruple systems if two of the bodies are appropriately replaced by a single body. In the case of 2+2 systems, we apply the stability criterion to the 1-3 and 2-3 orbit pairs; in the case of 3+1 systems, we apply the stability criterion to the 1-2 and 2-3 orbit pairs (see also \S\,\ref{sect:IC:orbits}). 
\item The system enters the semisecular regime. We use the criterion of \citet{2014ApJ...781...45A} applied to the 1-2 and 2-3 orbit pairs for 2+2 systems, and the 1-2 orbit pair for 3+1 systems (we do not consider the 2-3 orbit pair for 3+1 systems: for orbit pair 2-3 to enter the semisecular regime, $e_2$ would have to be very high, but the system would more likely become dynamically unstable before reaching such high $e_2$). Specifically, we assume that the semisecular regime is entered if
\begin{align}
\sqrt{1-e_\mathrm{in}} < 5 \pi \frac{m_\mathrm{out}}{m_\mathrm{in}} \left [ \frac{a_\mathrm{in}}{a_\mathrm{out} \left(1-e_\mathrm{out} \right )} \right ]^3,
\end{align}
where, for 2+2 systems and orbit 1, the inner and outer orbits are orbit 1 and 3, respectively, $m_\mathrm{in}=m_1+m_2$ and $m_\mathrm{out}=m_3+m_4$; for 2+2 systems and orbit 2, the inner and outer orbits are orbit 2 and 3, respectively, $m_\mathrm{in}=m_3+m_4$ and $m_\mathrm{out}=m_1+m_2$; for 3+1 systems and orbit 1, the inner and outer orbits are orbit 1 and 2, respectively, $m_\mathrm{in}=m_1+m_2$ and $m_\mathrm{out}=m_3$.
\item The age of the system exceeds 10 Gyr.
\end{enumerate}

We note that conditions (i) through (iii) are implemented as (instantaneous) root finding conditions within the set of ordinary differential equations (which are solved using the \textsc{CVODE} routine, \citealt{1996ComPh..10..138C}, which supports root finding). Therefore, there is no risk of the stopping conditions being missed in the simulations due to a finite number of output snapshots.

\section{Population synthesis setup}
\label{sect:IC}
In this section, we describe the methodology used to set up the systems for the population synthesis. A summary of our assumptions is given in the third column of Table\,\ref{table:IC}. We sample $N_\mathrm{MC}=10^4$ systems for the 2+2 and 3+1 systems, both with and without the effects of flybys (\S\,\ref{sect:meth:flybys}), giving a total of $4\times10^4$ systems.

\begin{figure}
\center
\includegraphics[scale = 0.45, trim = 10mm -5mm 0mm 10mm]{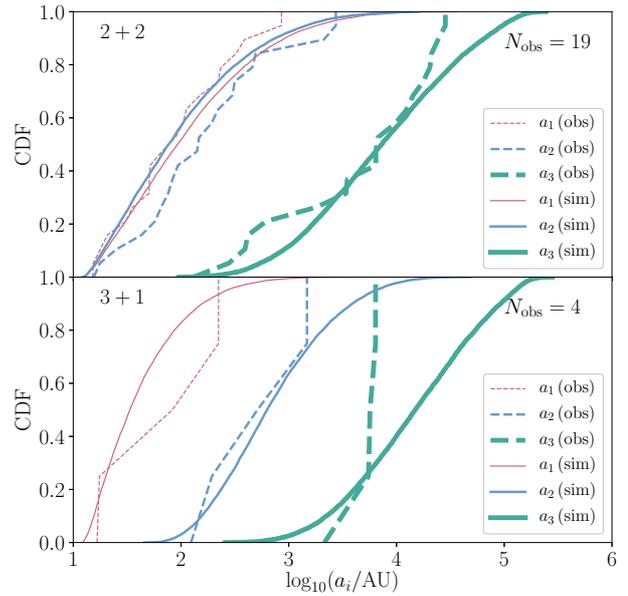}
\caption { Comparison of the semimajor axes sampled from the procedure described in \S\,\ref{sect:IC} (solid lines) to observational data satisfying similar requirements from the MSC (\citealt{1997A&AS..124...75T,2017arXiv171204750T}; dashed lines). The top (bottom) panel corresponds to the 2+2 (3+1) configuration. Orbits are indicated with different colours: red, blue and green for orbits 1 through 3. The number of systems in the MSC is indicated in the top right of each panel. }
\label{fig:IC_comp_sma}
\end{figure}

\begin{figure}
\center
\includegraphics[scale = 0.45, trim = 10mm -5mm 0mm 10mm]{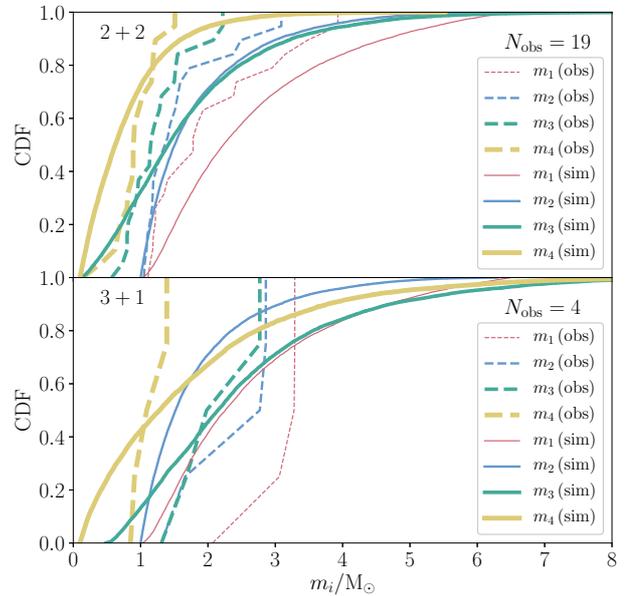}
\caption { Similar to \F\,\ref{fig:IC_comp_sma}, here comparing the distributions of the four masses. }
\label{fig:IC_comp_m}
\end{figure}

\subsection{Masses}
\label{sect:IC:masses}
Our focus is on systems in which the stars in at least one orbit are progenitors of CO WDs. Noting that zero-age MS stars of $1\,\msun$ will evolve to a CO WD in approximately 10 Gyr and that stars more massive than about $6.5\,\msun$ evolve to ONe WDs (assuming isolated stars and Solar metallicity), we therefore restrict to systems with $1<m_1/\msun<6.5$ and $m_2 > 1\, \msun$ (by definition, we set $m_2\leq m_1$). We assume a Kroupa initial mass function \citep{1993MNRAS.262..545K} for $m_1$, i.e., $\mathrm{d}N/\mathrm{d}m_1 \propto m_1^{-2.7}$ for $1<m_1/\msun<6.5$. The secondary mass, $m_2$, is sampled from a flat distribution of $q_1 = m_2/m_1$ \citep{2012Sci...337..444S,2013ARA&A..51..269D,2017ApJS..230...15M}, with $0.01<q_1\leq1$. 

Once $m_1$ and $m_2$ have been sampled for a system, the masses $m_3>0.1\,\msun$ and $m_4>0.1\,\msun$ are drawn depending on the configuration. For 2+2 systems, we sample a mass ratio $q_3 = (m_3+m_4)/(m_1+m_2)$ from a flat distribution with $0.01<q_3\leq1$, and another mass ratio $q_2 = m_4/m_3$, again assuming a flat distribution with $0.01<q_2\leq1$. This implies that the mass of binary 2 is correlated with that of binary 1. For 3+1 systems, we sample $q_2 = m_3/(m_1+m_2)$ from a flat distribution with $0.01<q_2\leq1$, and $q_3 = m_4/m_3$ from a flat distribution with $0.01<q_3\leq1$, again implying correlations of the masses of the outer stars with those of the inner stars.

\subsection{Orbits}
\label{sect:IC:orbits}
For both configurations, three orbital periods are drawn from a Gaussian distribution in $\mathrm{log}_{10}(P_{\mathrm{orb},i}/\mathrm{d})$, with a mean of 5.03 and standard deviation of 2.28, and $0<\mathrm{log}_{10}(P_{\mathrm{orb},i}/\mathrm{d})<10$ \citep{2010ApJS..190....1R}. The corresponding semimajor axes are computed according to the configuration using Kepler's law. In addition, three eccentricities are drawn from a Rayleigh distribution between 0.01 and 0.9 with an rms width of 0.33 \citep{2010ApJS..190....1R}. Subsequently, we impose stability criteria to ensure that the systems are dynamically stable using the criterion of \citeauthor{2001MNRAS.321..398M} (\citeyear{2001MNRAS.321..398M}; implicitly assuming that this also applies to quadruple systems), and conditions such that the stars would not evolve if they were part of isolated binaries. 

Specifically, let the criterion of \citet{2001MNRAS.321..398M}, as applied to a triple system with the inner and outer orbits indicated with `in' and `out', respectively, be denoted with $a_\mathrm{out}/a_\mathrm{in} > f_\mathrm{MA11}(m_\mathrm{in},m_\mathrm{out},e_\mathrm{out})$, where the latter function is given by
\begin{align}
f_\mathrm{MA11}(m_\mathrm{in},m_\mathrm{out},e_\mathrm{out}) \equiv \frac{2.8}{1-e_\mathrm{out}} \left [ \left (1+\frac{m_\mathrm{out}}{m_\mathrm{in}} \right ) \frac{1+e_\mathrm{out}}{\sqrt{1-e_\mathrm{out}}} \right ]^{2/5}.
\end{align}
For 2+2 systems, we impose
\begin{subequations}
\begin{align}
a_2/a_1 &>  f_\mathrm{MA11}(m_1+m_2,m_3+m_4,e_3); \\
a_3/a_2 &> f_\mathrm{MA11}(m_3+m_4,m_1+m_2,e_3); \\
a_1\left(1-e_1^2\right) &> 12 \, \au; \\
a_2\left(1-e_2^2\right) &> 12 \, \au,
\end{align}
\end{subequations}
whereas for 3+1 systems, we require
\begin{subequations}
\begin{align}
a_2/a_1 &>  f_\mathrm{MA11}(m_1+m_2,m_3,e_2); \\
a_3/a_2 &> f_\mathrm{MA11}(m_1+m_2+m_3,m_4,e_3); \\
a_1\left(1-e_1^2\right) &> 12 \, \au. 
\end{align}
\end{subequations}
The restrictions on the semilatus recti, $a_i\left(1-e_i^2\right)$, ensure that binaries 1 and 2 (for 2+2 systems), and binary 1 (for 3+1 systems) would not interact (i.e., would not be triggered into RLOF) if they were to evolve in isolation.

The initial orbital orientations are assumed to be random, i.e., for each orbit $i$, flat distributions are assumed for $\cos(i_i)$, $\omega_i$ and $\Omega_i$, where $i_i$, $\omega_i$ and $\Omega_i$ are the inclination, argument of periapsis, and longitude of the ascending node, respectively, of orbit $i$ (the orbital elements are defined with respect to an arbitrary fixed frame).

\subsection{Comparison to the Multiple Star Catalogue}
The statistics of orbital properties of quadruple star systems, especially when restricting to the systems of interest here, are poorly constrained. In particular, we find only a total of 23 systems satisfying the criteria given above in the Multiple Star Catalogue (MSC; \citealt{1997A&AS..124...75T,2017arXiv171204750T}). Nevertheless, we briefly compare our sampling methodology to the MSC in terms of the semimajor axis distributions (\F\,\ref{fig:IC_comp_sma}) and the mass distributions (\F\,\ref{fig:IC_comp_m}). In these two figures, the top (bottom) panel corresponds to the 2+2 (3+1) configuration. 

The distributions of the sampled semimajor axes (\F\,\ref{fig:IC_comp_sma}) are consistent with the MSC data, with the notable exception that the MSC data does not contain systems with large values of $a_3$, $a_3 \gtrsim 10^5\,\au$. Presumably, apart from the small number of systems in the observed sample, this is due to the difficulty of observing quadruple systems with very wide outermost orbits. The observed masses (\F\,\ref{fig:IC_comp_m}) tend to be confined to more narrow distributions compared to the sampled masses, but again the number of observed systems is low.

\section{Examples}
\label{sect:examples}
\begin{figure*}
\center
\includegraphics[scale = 0.65, trim = 15mm 10mm 0mm 0mm]{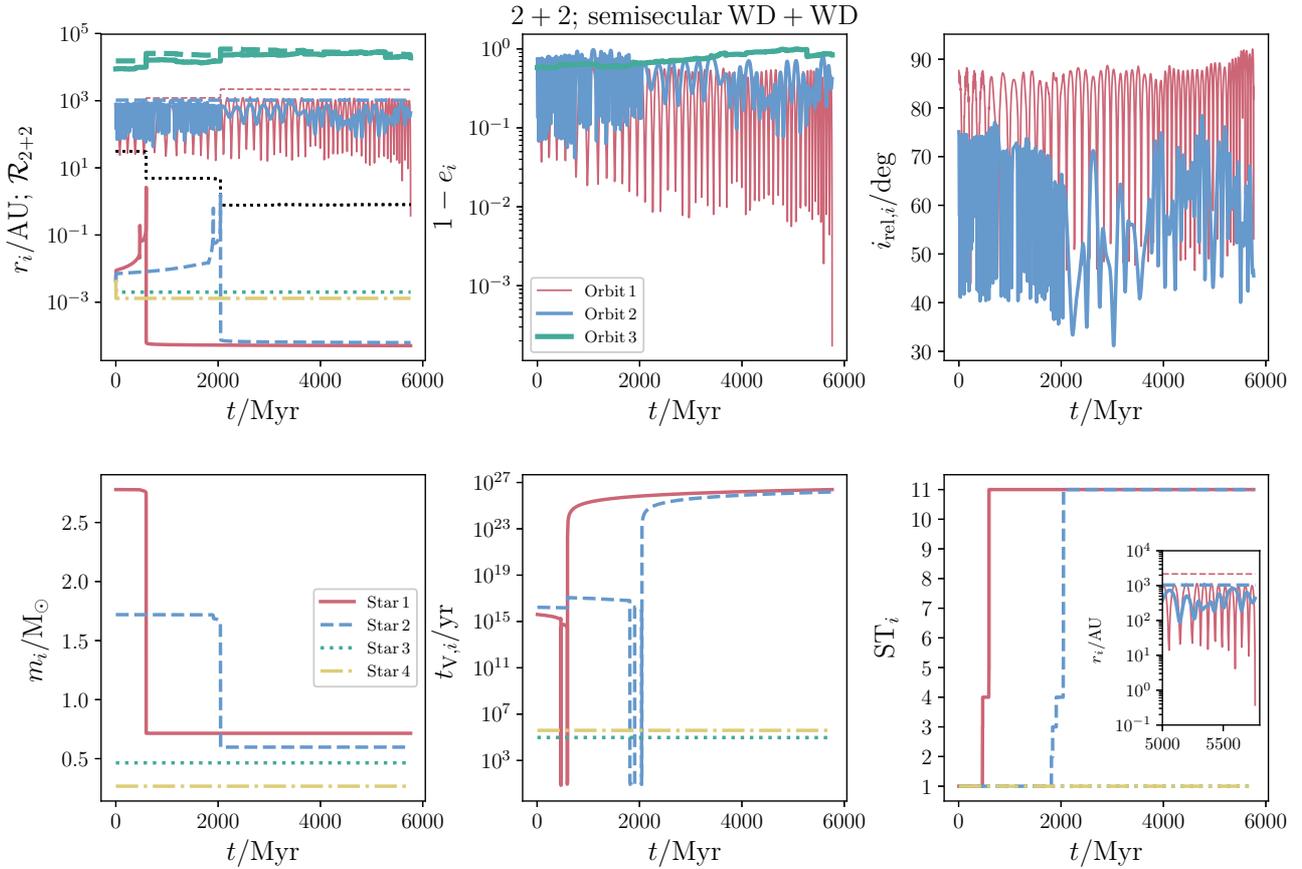}
\caption { Example evolution for a 2+2 system leading to a double CO WD system in the semisecular regime. Top-left panel: the semimajor axes (dashed lines), periapsis distances $a_i(1-e_i)$ (solid lines) and stellar radii (solid, dashed, dotted and dot-dashed lines for stars 1 through 4) as a function of time. For the orbits, red, blue and green correspond to orbits 1, 2 and 3, respectively; for the stars, red, blue, green and yellow correspond to stars 1, 2, 3 and 4, respectively. The black dotted line in the top-left panel shows the ratio of the LK time-scales $\mathcal{R}_{2+2}$ (equation~\ref{eq:R_2p2}). Top-middle panel: eccentricities; top-right panel: the relative inclinations of orbits 1 and 2 to their parent (i.e., $i_{13}$, red line, and $i_{23}$, blue line). Bottom-left panel: stellar masses; bottom-middle panel: viscous time-scales; bottom-right panel: stellar types (the same as in \citealt{2002MNRAS.329..897H}, see also the text in the beginning of \S\,\ref{sect:examples}). The inset in the bottom-right panel is a close-up of the top-left panel, i.e., showing the semimajor axis and periapsis distance evolution in detail near the end of the simulation. Refer to \S\,\ref{sect:examples:1} for a detailed description of the evolution. }
\label{fig:example1}
\end{figure*}

\begin{figure*}
\center
\includegraphics[scale = 0.65, trim = 15mm 10mm 0mm 0mm]{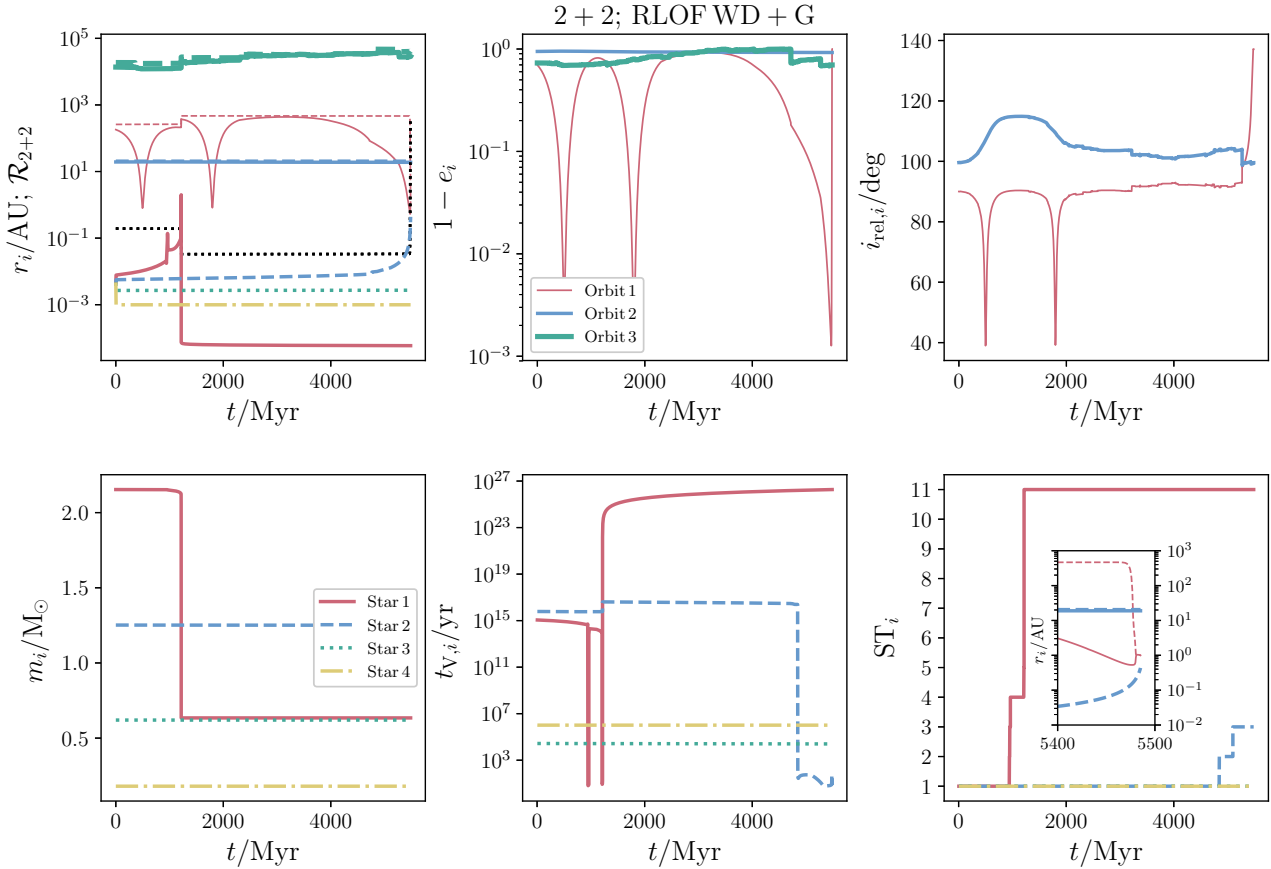}
\caption {Example of a 2+2 system in which the secondary star fills its Roche lobe as an RGB star when the primary has evolved to a CO WD. See the caption of \F\,\ref{fig:example1} for an explanation of the colours and line styles used, and \S\,\ref{sect:examples:2} for a detailed description of the evolution. }
\label{fig:example2}
\end{figure*}

\begin{figure*}
\center
\includegraphics[scale = 0.65, trim = 15mm 10mm 0mm 0mm]{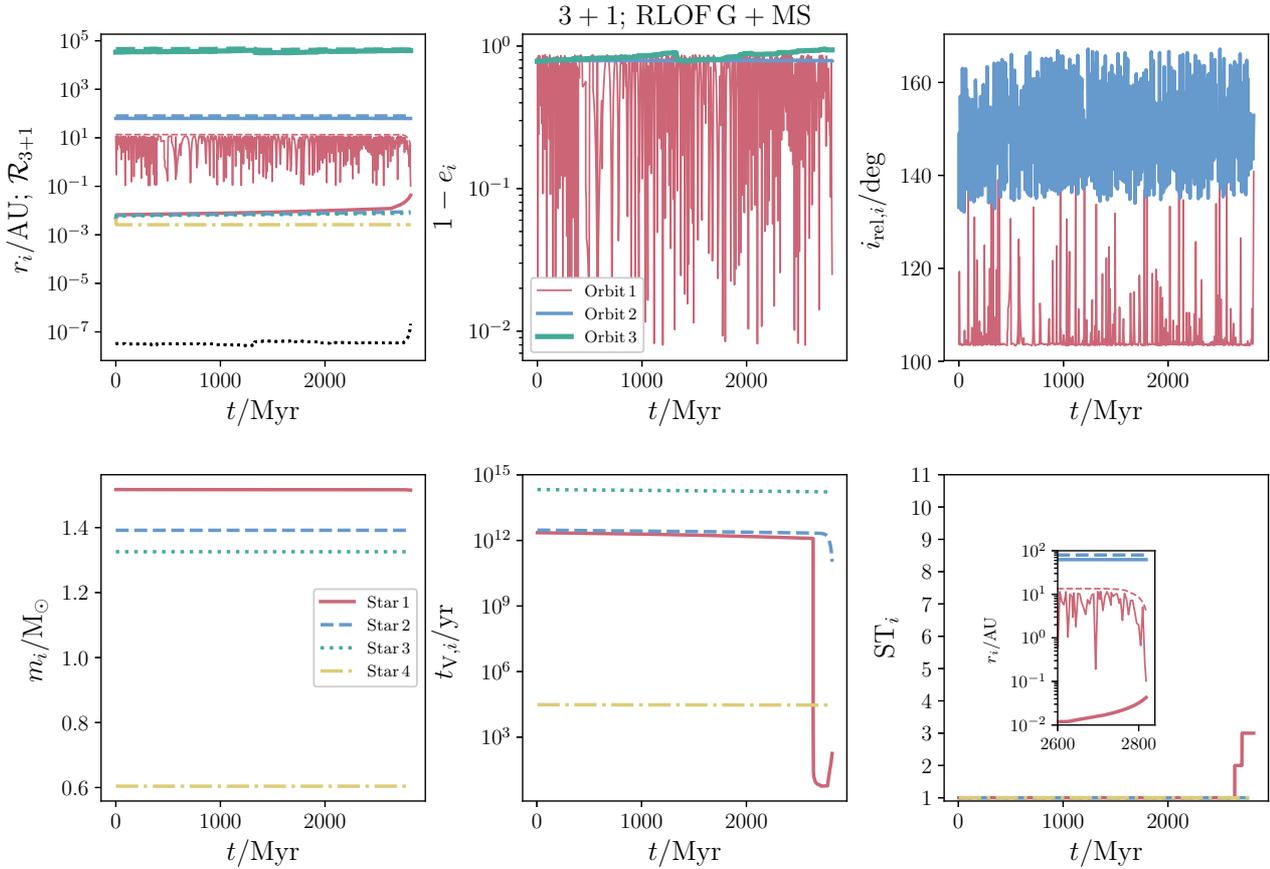}
\caption {Example of a 3+1 system in which the primary star fills its Roche lobe as an RGB star, with the secondary still on the MS. See the caption of \F\,\ref{fig:example1} for an explanation of the colours and line styles used, and \S\,\ref{sect:examples:3} for a detailed description of the evolution. }
\label{fig:example3}
\end{figure*}

\begin{figure*}
\center
\includegraphics[scale = 0.65, trim = 15mm 10mm 0mm 0mm]{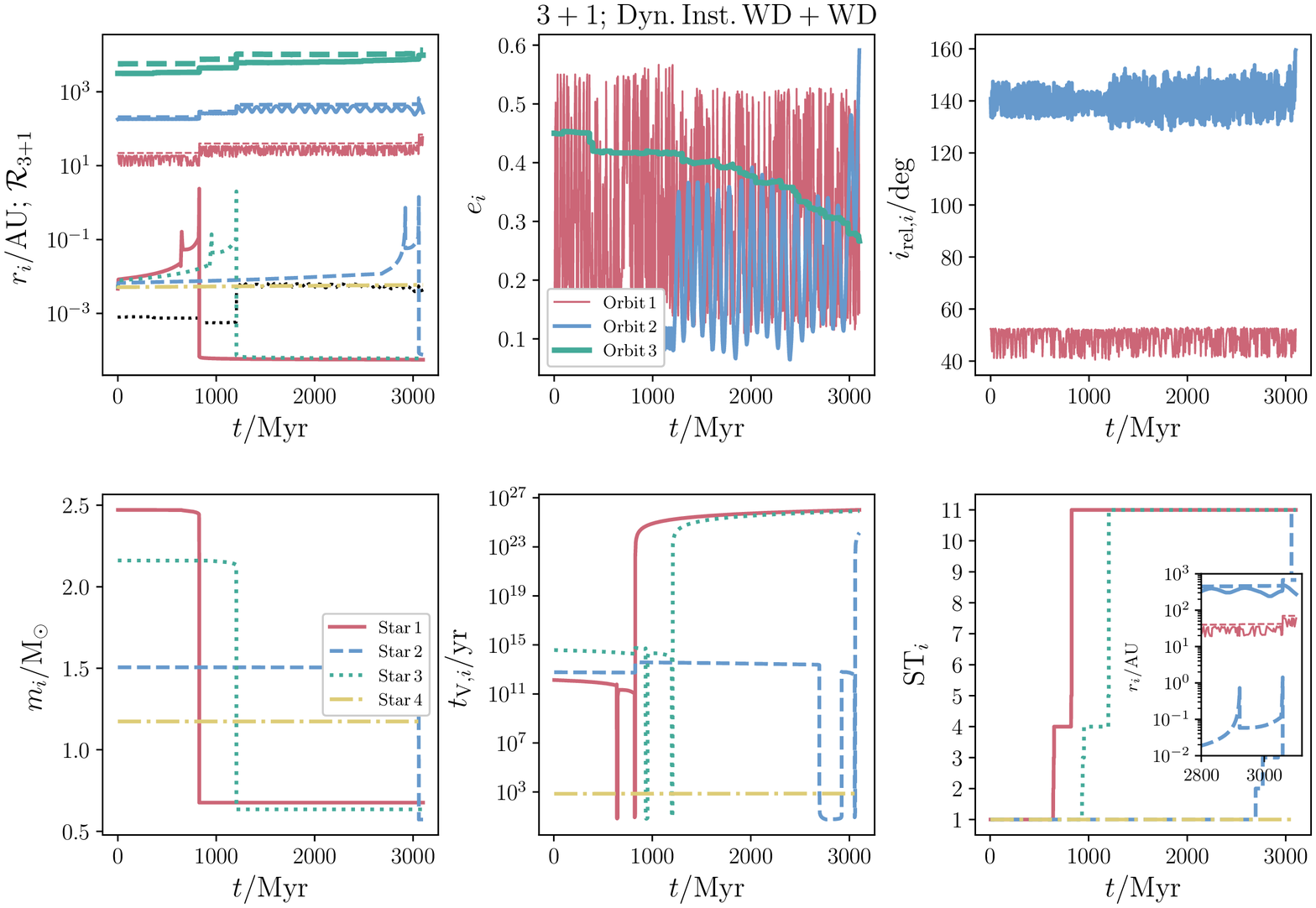}
\caption {Example of a 3+1 system in which a dynamical instability of orbit pair 1-2 occurs when stars 1, 2 and 3 are CO WDs. See the caption of \F\,\ref{fig:example1} for an explanation of the colours and line styles used, and \S\,\ref{sect:examples:4} for a detailed description of the evolution. }
\label{fig:example4}
\end{figure*}

Here, we illustrate several evolutionary pathways found in the population synthesis calculations by giving a number of examples in Figs\,\ref{fig:example1}, \ref{fig:example2}, \ref{fig:example3}, and \ref{fig:example4}. For each example we plot, as a function of time  in the top-left panels, the semimajor axes (dashed lines), periapsis distances $a_i(1-e_i)$ (solid lines) and stellar radii (solid, dashed, dotted and dot-dashed lines for stars 1 through 4). Also in the top-left panels, we show with the black dotted lines ratios of LK time-scales, i.e.,
\begin{subequations}
\begin{align}
\label{eq:R_2p2}
\mathcal{R}_{2+2} &\equiv \frac{t_\mathrm{LK,13}}{t_\mathrm{LK,23}} \simeq \left ( \frac{a_2}{a_1} \right)^{3/2} \left ( \frac{m_1+m_2}{m_3+m_4} \right )^{3/2}; \\
\label{eq:R_3p1}
\mathcal{R}_{3+1} &\equiv \frac{t_\mathrm{LK,12}}{t_\mathrm{LK,23}} \simeq \left ( \frac{a_2^3}{a_1 a_3^2} \right)^{3/2} \left ( \frac{m_1+m_2}{m_1+m_2 + m_3} \right )^{1/2} \frac{m_4}{m_3} \left ( \frac{1-e_2^2}{1-e_3^2} \right)^{3/2}.
\end{align}
\end{subequations}
When these ratios are close to unity (roughly speaking, within an order of magnitude), secularly chaotic behaviour and particularly high eccentricities are to be expected \citep{2017MNRAS.470.1657H}. The top-middle panels show the eccentricities, and the top-right panels show the relative inclinations of orbits 1 and 2 to their parent (i.e., $i_{13}$ and $i_{23}$ for the 2+2 configuration, and $i_{12}$ and $i_{23}$ for the 3+1 configuration). The bottom-left panels show the stellar masses, the bottom-middle panels show the viscous time-scales, and the bottom-right panels show the stellar types (the same as in \citealt{2002MNRAS.329..897H}). The stellar types of relevance here are 1 -- MS, 2 -- Hertzsprung gap (HG), 3 -- red giant branch (RGB), 4 -- core helium burning (CHeB), 5 -- early asymptotic giant branch (AGB), 6 -- thermally pulsing AGB, and 11 -- CO WD. Also shown in the insets in the bottom-right panels are close-ups of the top-left panel, i.e., displaying the semimajor axis and periapsis distance evolution in detail near the end of the simulation.

\subsection{2+2: WD collision in the semisecular regime}
\label{sect:examples:1}
In \F\,\ref{fig:example1}, both orbits 1 and 2 are initially inclined with respect to orbit 3. Consequently, they are excited in their eccentricity during the MS, although not sufficiently to trigger RLOF or tidal evolution (note that the viscous time-scales on the MS are extremely long). When the primary star evolves to a CO WD after $\approx 500\,\mathrm{Myr}$, orbit 1 expands, bringing its semimajor axis closer to that of orbit 2, and the total mass contained in orbit 1 becomes more similar to that of orbit 2 implying that $\mathcal{R}_{2+2}$ (equation~\ref{eq:R_2p2}) becomes closer to unity. The maximum eccentricity reached in orbit 1 indeed increases after the primary star has evolved to a WD. Later, after $\approx 2000\,\mathrm{Myr}$, the secondary star evolves to a CO WD, and the ratio $\mathcal{R}_{2+2}$ approaches unity even more closely. The maximum eccentricity in orbit 1 continues to increase over time; close to $6000\,\mathrm{Myr}$, orbit 1 switches from prograde to retrograde and vice versa with respect to orbit 3. The eccentricity of orbit 1 then peaks to $\sim 1-10^{-4}$, and orbit 1 enters the semisecular regime after which the simulation is stopped. The inner orbit consists of two CO WDs at this stage, and a collision (and hence, potentially a SNe Ia) can be expected after a certain delay time (see also \S\,\ref{sect:results:col} below).

\subsection{2+2: RLOF of a giant star with a WD companion}
\label{sect:examples:2}
In the previous example, stellar evolution aided in the process of producing a system with (eventually) colliding WDs by gradually fine-tuning $\mathcal{R}_{2+2}$ to unity (thus avoiding collisions on the MS). In \F\,\ref{fig:example2}, we show a counterexample in which, in a not very dissimilar 2+2 system, stellar evolution disrupts the above formation process of a WD collision system by triggering RLOF. In particular, orbits 1 and 2 are highly inclined with respect to orbit 3, but the LK time-scales are too long to induce a merger on the MS. When the primary star evolves to a CO WD at $\approx 1600\,\mathrm{Myr}$, orbit 1 happens to be near a minimum eccentricity, and it can accommodate the primary when it is an AGB star (with radius $\sim 1\, \au$) without triggering tidal evolution or RLOF. However, when the secondary evolves to an RGB star at $\approx 6000 \, \mathrm{Myr}$, orbit 1 is close to a maximum eccentricity, and, combined with a much larger radius and higher tidal dissipation efficiency of the secondary star (viscous time-scale of $\sim 10^1\,\mathrm{yr}$) as it evolves off of the MS, tidal dissipation shrinks the semimajor axis of orbit 1 from $\sim 5\times 10^2\,\au$ to $\sim 1 \,\au$, and circularises it. Subsequently, orbit 1 can no longer accommodate the secondary star with its radius approaching $1\,\au$, and the secondary star fills its Roche lobe.

The further evolution is not modelled here, but one can expect CE evolution to occur given that the primary is a WD, finally producing a tight double WD binary in which further secular evolution is quenched due to relativistic precession. This channel, analogous to that found previously for triple systems \citep{2013MNRAS.430.2262H}, might lead to a double CO WD merger provided that the GW merger time-scale is sufficiently short, but its orbit would be circular at the time of merger, and be dynamically distinct from colliding WDs.

\subsection{3+1: RLOF of a giant star with a MS companion}
\label{sect:examples:3}
The last two examples illustrate the evolution of 3+1 systems. In \F\,\ref{fig:example3}, the initial ratio $\mathcal{R}_{3+1}\sim 10^{-7}$ is small and far from unity, indicating that the fourth star can be neglected from a secular dynamical point of view \citep{2015MNRAS.449.4221H}. Nevertheless, orbits 1 and 2 are initially highly inclined, and the eccentricity of orbit 1 is excited to $\sim 1-10^{-2}$. The latter is not sufficient, however, to trigger interaction during the MS stage of the primary star. Later, as the primary star evolves to an RGB star, the short-period high-amplitude eccentricity oscillations, coupled with the increased radius and higher tidal dissipation efficiency of the primary star, drive orbit 1 to shrink due to tidal dissipation and circularise. RLOF is then triggered after $\approx 2800\, \mathrm{Myr}$.

\subsection{3+1: Dynamical instability with three WDs}
\label{sect:examples:4}
In the last example (\F\,\ref{fig:example4}), we illustrate how a 3+1 system with three CO WDs could be driven to dynamical instability, potentially resulting in collisions. Orbits 1 and 2 are initially mildly inclined ($i_\mathrm{12,i} \approx 50^\circ$) and the inner triple is again decoupled from the fourth star ($\mathcal{R}_{3+1}\sim 10^{-3}$), yielding maximum eccentricities of $\sim 0.6$ in orbit 1. These maximum eccentricities are not high enough to trigger interaction during the MS, or when the primary star evolves to a CO WD. At $\sim 1200\,\mathrm{Myr}$, the third star evolves to a CO WD; due to mass loss and the effects on the orbits, orbits 2 and 3 are brought into a more strongly-interacting LK regime (i.e., the ratio $a_3/a_2$ decreases and $m_4/m_3$ increases), and $\mathcal{R}_{3+1}$ (equation~\ref{eq:R_3p1}) increases. However, $\mathcal{R}_{3+1}$ is still $\lesssim 10^{-2}$ and the inner triple remains decoupled from the fourth star. After $\approx 3000\,\mathrm{Myr}$, the secondary star evolves to a CO WD. The subsequent relative orbital expansion (i.e., $a_2/a_1$ decreases), then drives dynamical instability, similar to triple systems \citep{2012ApJ...760...99P}.

\section{Results}
\label{sect:results}
Our main results are presented in this section. We first discuss the outcome fractions of the various channels in \S\,\ref{sect:results:frac}. In \S\,\ref{sect:results:orb}, we consider orbital properties for the various outcomes and orbital evolution, and in \S\,\ref{sect:results:time} we investigate the stopping times. We present results related to WD collisions, and potential SNe Ia progenitors, in \S\,\ref{sect:results:col}.

\subsection{Outcome fractions}
\label{sect:results:frac}
The results from the population synthesis calculations can be summarised with Table\,\ref{table:fractions}, which gives the fractions of several channels of interest. The fractions are determined from $N_\mathrm{MC}=10^4$ simulations, and are given for the 2+2 and 3+1 configurations after a random time (indicated with the columns $t_x$) or after 10 Gyr of evolution (if applicable; otherwise, the end time is the stopping condition time), and with and without taking into account the effects of flybys (\S\,\ref{sect:meth:flybys}). The outcomes are based on the stopping conditions adopted in the simulations (\S\,\ref{sect:meth:sc}). 

For both the 2+2 and 3+1 configurations, the most likely outcome is `no interaction', i.e., the system reaches an age of 10 Gyr without triggering other stopping conditions. The orbits may still have evolved (in particular, expanded due to mass loss; see also \S\,\ref{sect:results:orb}). RLOF of one of the four stars is triggered in about $\approx 0.2$ of systems, with RLOF of the primary and tertiary star most common, and most likely during the MS or giant stages. Note that we did not check for RLOF for stars 3 and 4 in the 3+1 configuration (given their wide orbits, this would be highly unlikely). A dynamical instability can be triggered in orbit 1 (i.e., orbit 1 becomes unstable with respect to its parent orbit), or orbit 2. The corresponding rows in Table\,\ref{table:fractions} give the total fraction of dynamically unstable systems, and the branching over several combinations of the stars in the inner orbit of the instability pair  (e.g., the `MS+MS' row in the dynamical instability orbit 1 section gives the fraction of systems in which orbit 1 became dynamically unstable with respect to its parent, and the primary and secondary stars were MS stars at the time of instability). Similarly, we give the fractions of systems in which the semisecular regime was entered (orbits 1 and 2), and make a distinction between the evolutionary stages of the stars involved. Note that we did not check for the semisecular regime in orbit 2 for the 3+1 configuration. Lastly, for WDs we did not check for RLOF in the simulations but collisions instead (see \S\,\ref{sect:meth:sc}). Such collisions, induced by purely secular evolution, do occur in the simulations, but are extremely rare. In particular, the WD+WD secular collision fraction is $\approx 0.001$ for orbit 1 and both configurations. Only 3 secular WD collisions occurred in orbit 2 for the 2+2 configuration with flybys included (the corresponding fraction in Table\,\ref{table:fractions} is rounded to zero).

The no-interaction fractions are significantly lower for the 3+1 systems compared to the 2+2 systems ($\approx 0.4$ vs. $\approx 0.6$). This can be ascribed to the typically higher capacity of 3+1 systems to induce high eccentricities in the innermost orbit (orbit 1), since in 3+1 systems the eccentricity of orbit 2 can be excited efficiently due to LK oscillations with orbit 3 giving a shorter LK time-scale $t_\mathrm{LK,12} \propto (1-e_2^2)^{3/2}$, whereas for 2+2 systems the eccentricity of the parent orbit of orbit 1 is only affected by high-order (and hence small) secular terms (and flybys, but these apply predominantly to orbit 3 in both 2+2 and 3+1 systems). The RLOF fractions for stars 1 and 2 (in orbit 1) are indeed significantly higher for 3+1 systems (in total $\approx 0.2$) compared to 2+2 systems (in total $\approx 0.07$), and the semisecular fractions for orbit 1 are also larger for 3+1 systems compared to 2+2 systems. In addition, due to the nested nature of the orbits in the 3+1 configuration, the fraction of dynamically unstable systems (in particular for orbit 1) is significantly higher in 3+1 systems ($\approx 0.25$ vs. $0.03$ for orbit 1). 

Flybys have a relatively small effect on the fractions, changing them by no more than a few per cent. As expected, flybys typically tend to decrease the fraction of non-interacting systems and increase the fraction of the interacting systems (in particular, dynamical instability and semisecular regime). Although the effect of flybys is only a few percent on the fractions of the main channels, interestingly, for the WD+WD dynamical instability and semisecular outcomes (orbit 1), the effect can be relatively large. For example, for 2+2 systems the WD+WD dynamical instability fraction for orbit 1 after 10 Gyr increases from 0.003 to 0.010. We will evaluate below in \S\,\ref{sect:results:col} whether this increase is interesting in terms of the SNe Ia rates.

\definecolor{Gray}{gray}{0.9}
\begin{table*}
\begin{tabular}{lcccccccc}
\toprule
& \multicolumn{8}{c}{Fraction} \\
& \multicolumn{4}{c}{$2+2$} &\multicolumn{4}{c}{$3+1$} \\
& \multicolumn{2}{c}{$t_x$} & \multicolumn{2}{c}{10 Gyr} & \multicolumn{2}{c}{$t_x$} & \multicolumn{2}{c}{10 Gyr} \\
& NF & F & NF & F & NF & F & NF & F \\
\midrule
No Interaction & 0.656 & 0.645 & 0.635 & 0.616 & 0.458 & 0.443 & 0.428 & 0.398  \\
\midrule
RLOF $\star$1 & 0.048 & 0.048 & 0.050 & 0.050 & 0.179 & 0.184 & 0.190 & 0.196  \\
\quad MS & 0.012 & 0.015 & 0.012 & 0.016 & 0.036 & 0.040 & 0.036 & 0.040  \\
\quad G & 0.032 & 0.030 & 0.033 & 0.032 & 0.129 & 0.131 & 0.138 & 0.140  \\
\quad CHeB & 0.004 & 0.002 & 0.005 & 0.003 & 0.014 & 0.014 & 0.016 & 0.016  \\
\midrule
RLOF $\star$2 & 0.020 & 0.018 & 0.022 & 0.021 & 0.038 & 0.035 & 0.041 & 0.038  \\
\quad MS & 0.016 & 0.014 & 0.016 & 0.014 & 0.030 & 0.028 & 0.031 & 0.029  \\
\quad G & 0.004 & 0.003 & 0.005 & 0.005 & 0.006 & 0.006 & 0.008 & 0.008  \\
\quad CHeB & 0.001 & 0.001 & 0.001 & 0.002 & 0.001 & 0.001 & 0.002 & 0.001  \\
\midrule
RLOF $\star$3 & 0.069 & 0.085 & 0.075 & 0.091 & --- & --- & --- & ---  \\
\quad MS & 0.052 & 0.068 & 0.057 & 0.071 & --- & --- & --- & ---  \\
\quad G & 0.014 & 0.015 & 0.015 & 0.017 & --- & --- & --- & ---  \\
\quad CHeB & 0.002 & 0.002 & 0.003 & 0.003 & --- & --- & --- & ---  \\
\midrule
RLOF $\star$4 & 0.037 & 0.026 & 0.039 & 0.029 & --- & --- & --- & ---  \\
\quad MS & 0.036 & 0.026 & 0.038 & 0.027 & --- & --- & --- & ---  \\
\quad G & 0.000 & 0.001 & 0.001 & 0.001 & --- & --- & --- & ---  \\
\quad CHeB & 0.000 & 0.000 & 0.000 & 0.000 & --- & --- & --- & ---  \\
\midrule
Dynamical Instability Orbit 1 & 0.032 & 0.034 & 0.035 & 0.040 & 0.238 & 0.243 & 0.250 & 0.262  \\
\quad MS+MS & 0.001 & 0.002 & 0.001 & 0.002 & 0.150 & 0.154 & 0.151 & 0.155  \\
\quad G+MS & 0.011 & 0.010 & 0.011 & 0.010 & 0.019 & 0.019 & 0.020 & 0.020  \\
\quad WD+MS & 0.002 & 0.003 & 0.002 & 0.003 & 0.027 & 0.029 & 0.029 & 0.032  \\
\quad WD+G & 0.015 & 0.012 & 0.017 & 0.013 & 0.020 & 0.018 & 0.024 & 0.021  \\
\quad WD+WD & 0.002 & 0.005 & 0.003 & 0.010 & 0.013 & 0.016 & 0.018 & 0.027  \\
\midrule
Dynamical Instability Orbit 2 & 0.015 & 0.015 & 0.015 & 0.017 & 0.026 & 0.031 & 0.027 & 0.040  \\
\quad MS+MS & 0.000 & 0.001 & 0.000 & 0.002 & 0.010 & 0.016 & 0.010 & 0.020  \\
\quad G+MS & 0.012 & 0.011 & 0.012 & 0.011 & 0.005 & 0.004 & 0.005 & 0.004  \\
\quad WD+MS & 0.002 & 0.003 & 0.002 & 0.003 & 0.009 & 0.007 & 0.009 & 0.010  \\
\quad WD+G & 0.001 & 0.000 & 0.001 & 0.000 & 0.000 & 0.000 & 0.000 & 0.001  \\
\quad WD+WD & 0.000 & 0.000 & 0.000 & 0.001 & 0.000 & 0.001 & 0.000 & 0.003  \\
\midrule
Semisecular Orbit 1 & 0.035 & 0.038 & 0.038 & 0.043 & 0.061 & 0.063 & 0.063 & 0.065  \\
\quad MS+MS & 0.021 & 0.023 & 0.021 & 0.023 & 0.055 & 0.056 & 0.055 & 0.056  \\
\quad G+MS & 0.001 & 0.000 & 0.001 & 0.000 & 0.001 & 0.001 & 0.001 & 0.001  \\
\quad WD+MS & 0.007 & 0.008 & 0.008 & 0.009 & 0.003 & 0.003 & 0.003 & 0.003  \\
\quad WD+G & 0.001 & 0.001 & 0.001 & 0.001 & 0.000 & 0.000 & 0.000 & 0.000  \\
\quad WD+WD & 0.004 & 0.005 & 0.007 & 0.010 & 0.002 & 0.003 & 0.003 & 0.005  \\
\midrule
Semisecular Orbit 2 & 0.089 & 0.089 & 0.090 & 0.092 & --- & --- & --- & ---  \\
\quad MS+MS & 0.082 & 0.083 & 0.083 & 0.084 & --- & --- & --- & ---  \\
\quad G+MS & 0.001 & 0.001 & 0.001 & 0.002 & --- & --- & --- & ---  \\
\quad WD+MS & 0.005 & 0.004 & 0.005 & 0.005 & --- & --- & --- & ---  \\
\quad WD+G & 0.000 & 0.000 & 0.000 & 0.000 & --- & --- & --- & ---  \\
\quad WD+WD & 0.000 & 0.001 & 0.001 & 0.001 & --- & --- & --- & ---  \\
\midrule
Secular Collision Orbit 1 & 0.000 & 0.001 & 0.001 & 0.001 & 0.000 & 0.000 & 0.001 & 0.001  \\
\quad WD+WD & 0.000 & 0.001 & 0.001 & 0.001 & 0.000 & 0.000 & 0.001 & 0.001  \\
\midrule
Secular Collision Orbit 2 & 0.000 & 0.000 & 0.000 & 0.000 & --- & --- & --- & ---  \\
\quad WD+WD & 0.000 & 0.000 & 0.000 & 0.000 & --- & --- & --- & ---  \\
\bottomrule
\end{tabular}
\caption{ Fractions of outcomes from the population synthesis calculations. Columns 2-5: 2+2 systems; columns 6-9: 3+1 systems. Fractions are shown after a random time (indicated with the columns $t_x$) or after 10 Gyr of evolution (if applicable; otherwise, the end time is the stopping condition time). Results are shown without the inclusion of flybys (`NF'), or with inclusion (`F'). The fractions in each column are obtained from $N_\mathrm{MC}=10^4$ simulations, and are rounded to three decimal places. For the dynamical instability, semisecular and secular collision outcomes, we give the total fractions, and the fractions making a distinction between the types of stars in the orbit indicated. Here, `G' means giant(like), i.e., an HG, RGB or AGB star. For example, the dynamical instability orbit 2 `G+MS' row shows the fractions of systems in which dynamical instability occurred in orbit 2, and for which the stars in orbit 2 at the time of dynamical instability consisted of a giant and MS star. Note that some outcomes do not apply in the simulations of the 3+1 configuration (indicated with `---'): RLOF for stars 3 and 4, and the semisecular regime and secular collision for orbit 2. }
\label{table:fractions}
\end{table*}

\subsection{Orbital properties and evolution}
\label{sect:results:orb}
\begin{figure*}
\center
\includegraphics[scale = 0.45, trim = 10mm -5mm 0mm 10mm]{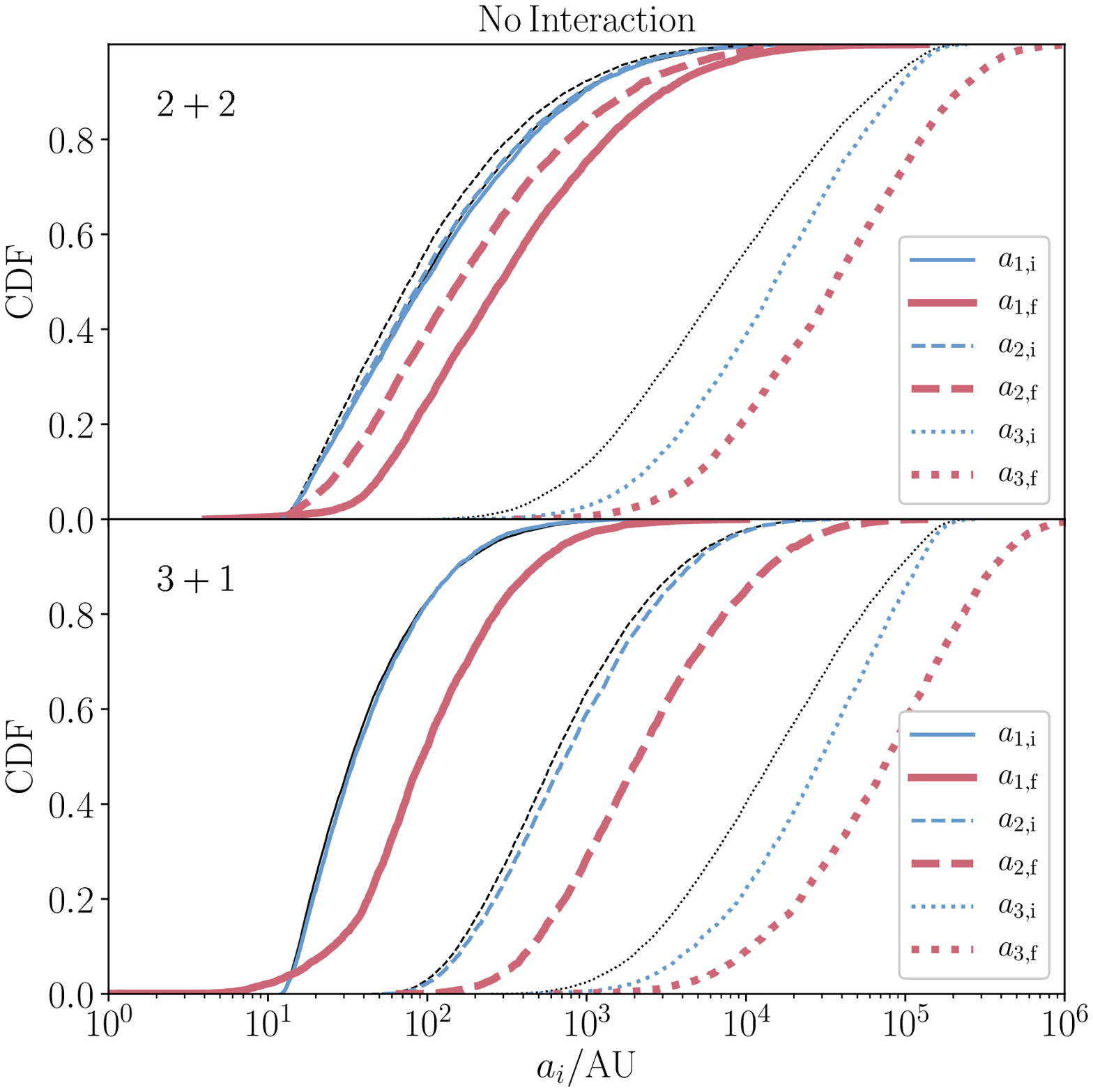}
\includegraphics[scale = 0.45, trim = 10mm -5mm 0mm 10mm]{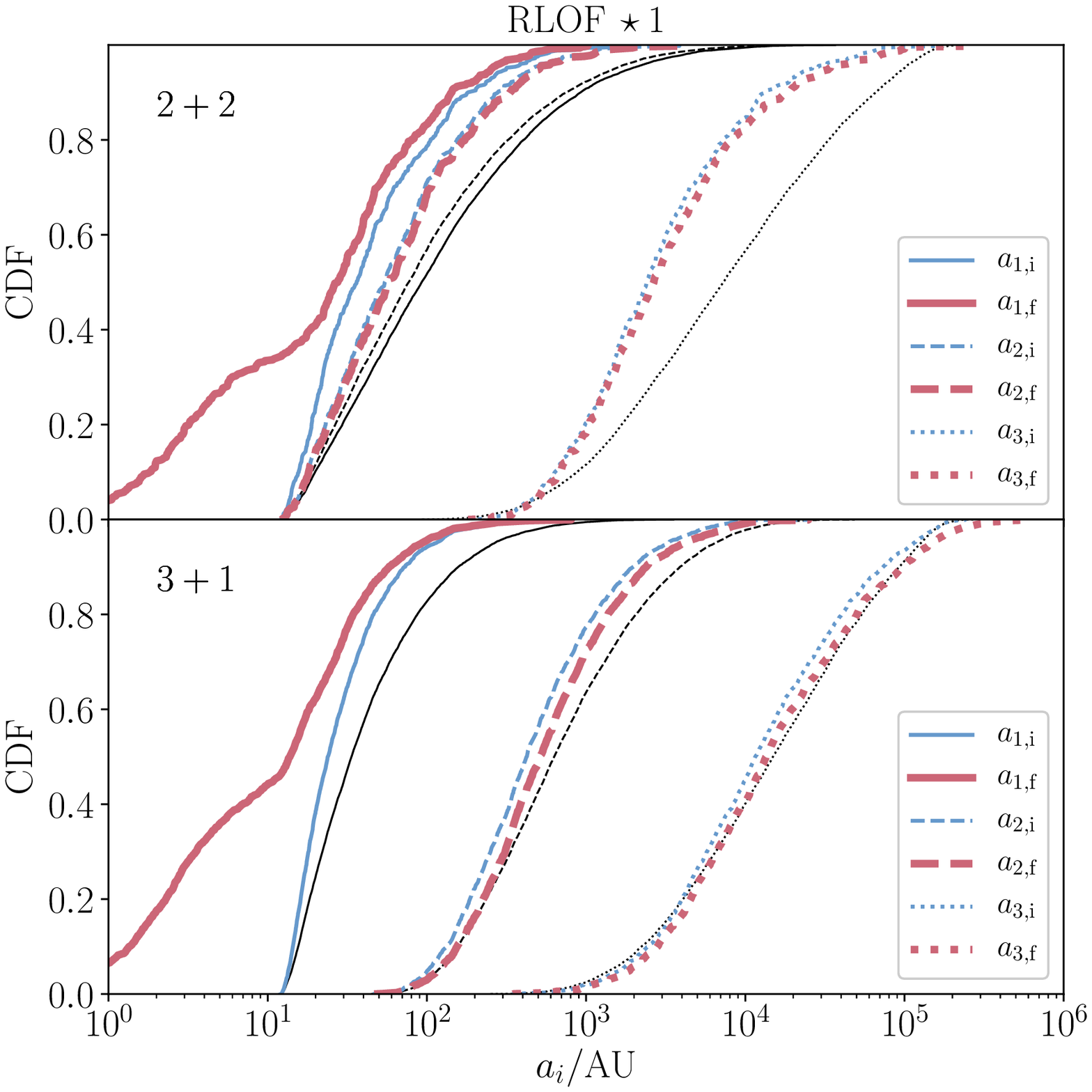}
\includegraphics[scale = 0.45, trim = 10mm -5mm 0mm 10mm]{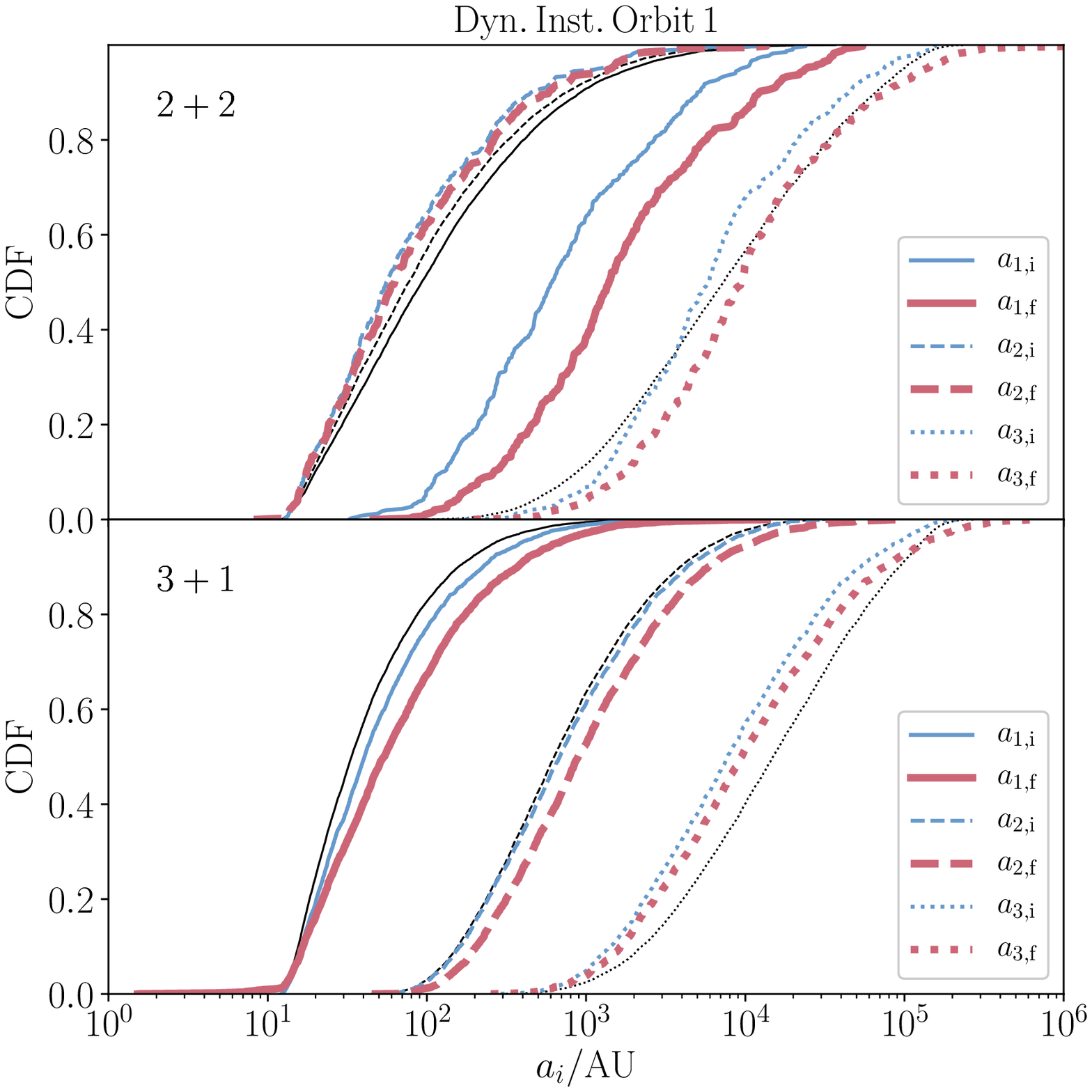}
\includegraphics[scale = 0.45, trim = 10mm -5mm 0mm 10mm]{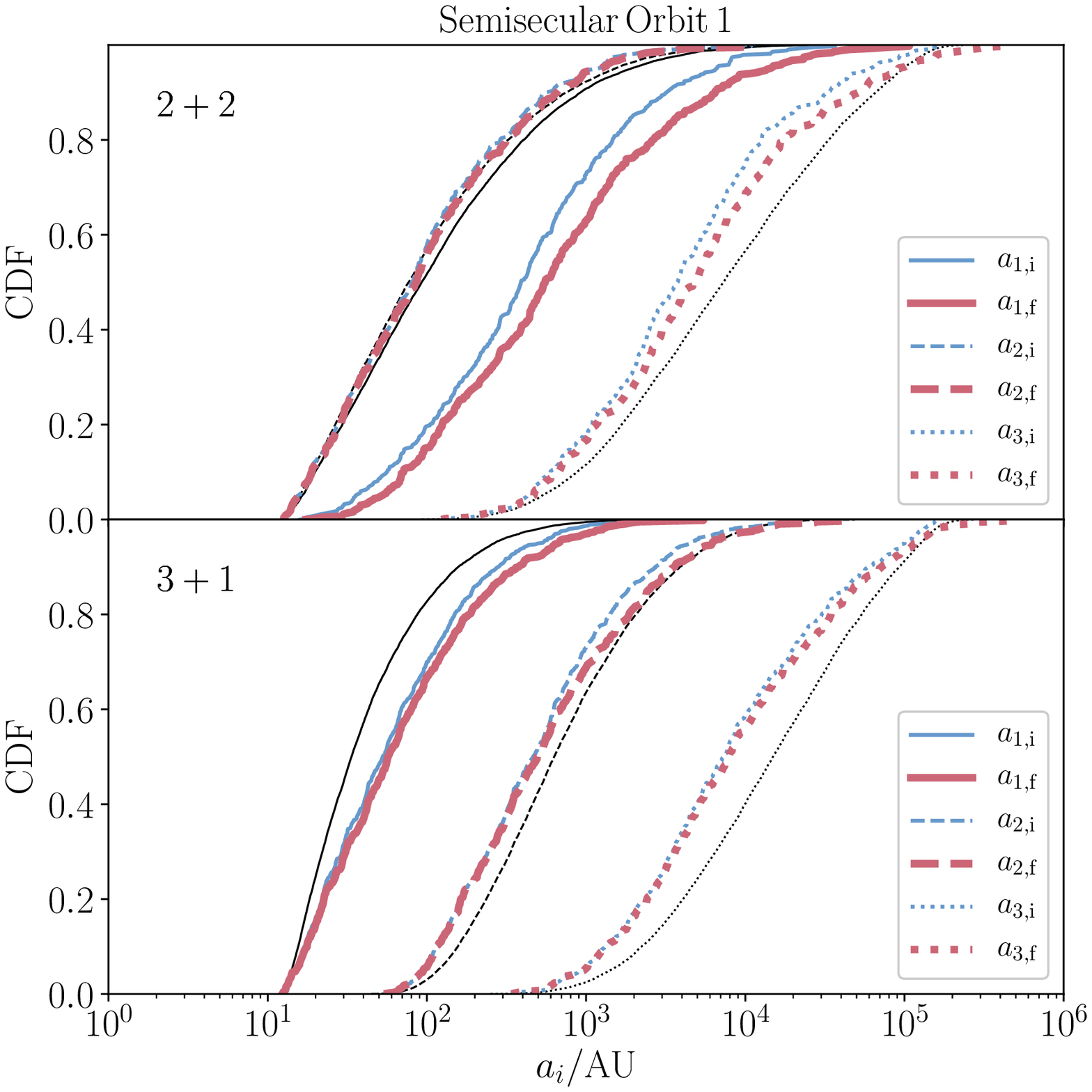}
\caption { Cumulative distributions of the semimajor axes for four different outcomes in the population synthesis simulations, including flybys. Each set of two panels (top: 2+2 configuration; bottom: 3+1 configuration) corresponds to a different outcome, indicated in the top panel. Shown are distributions of the initial values (blue lines), and values at the end of the simulation (red lines). The thin black lines show the initial distributions for {\it all} systems (i.e., not making a distinction of the outcome), and are the same in each set of two panels. The line styles indicate the different orbits --- solid: orbit 1; dashed: orbit 2; dotted: orbit 3. }
\label{fig:pop_syn_sma}
\end{figure*}

\begin{figure*}
\center
\includegraphics[scale = 0.45, trim = 10mm -5mm 0mm 10mm]{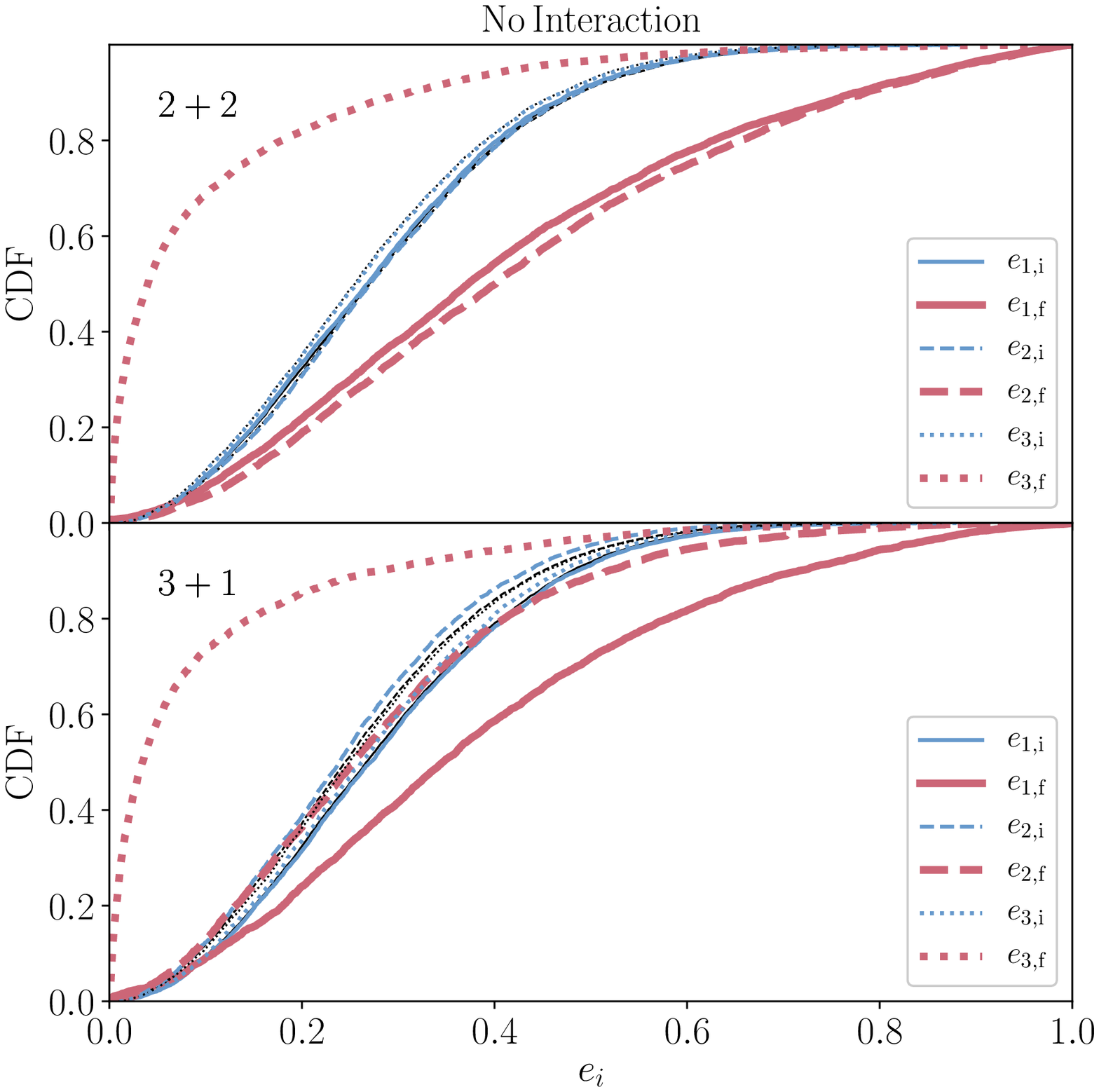}
\includegraphics[scale = 0.45, trim = 10mm -5mm 0mm 10mm]{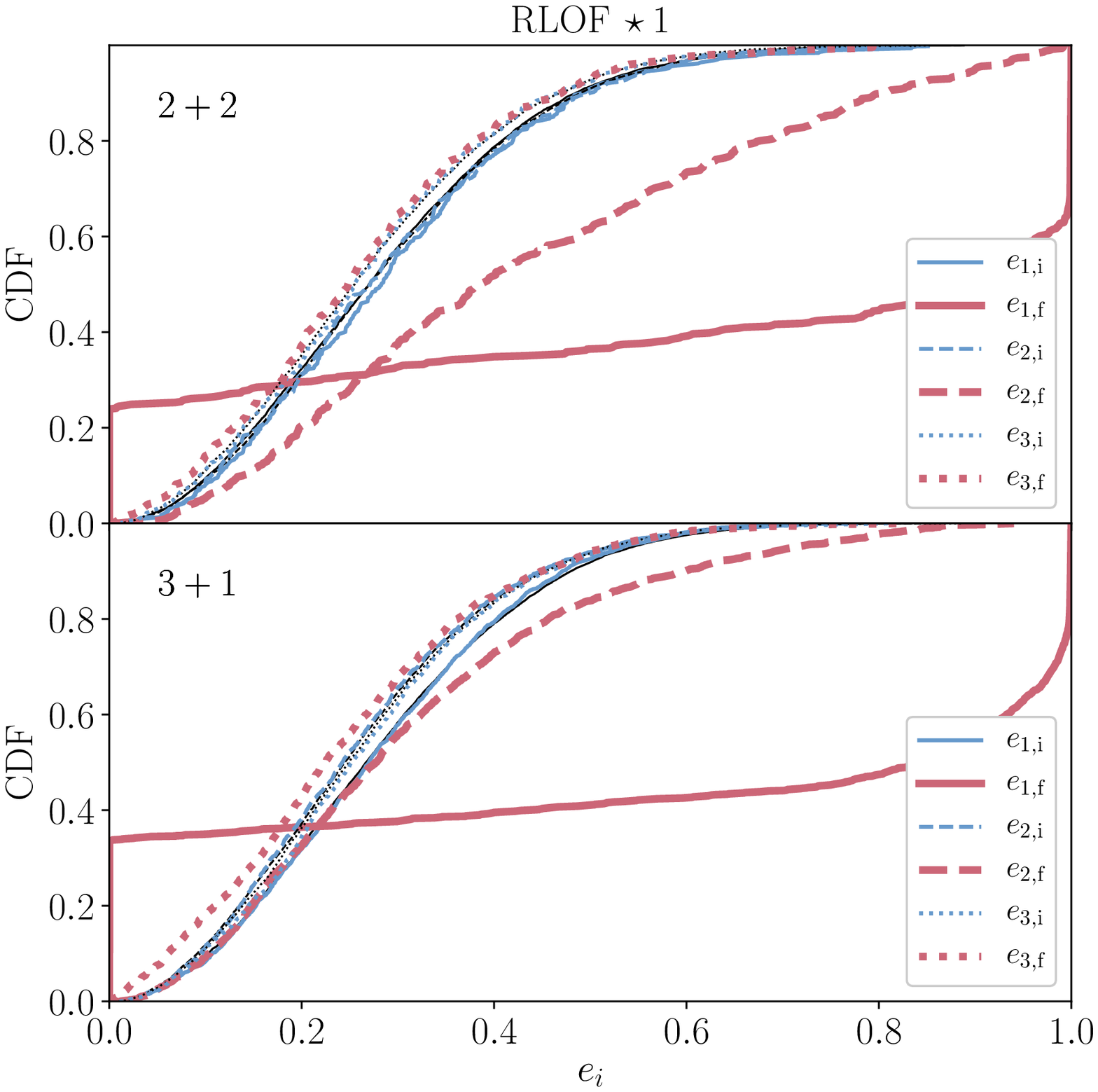}
\includegraphics[scale = 0.45, trim = 10mm -5mm 0mm 10mm]{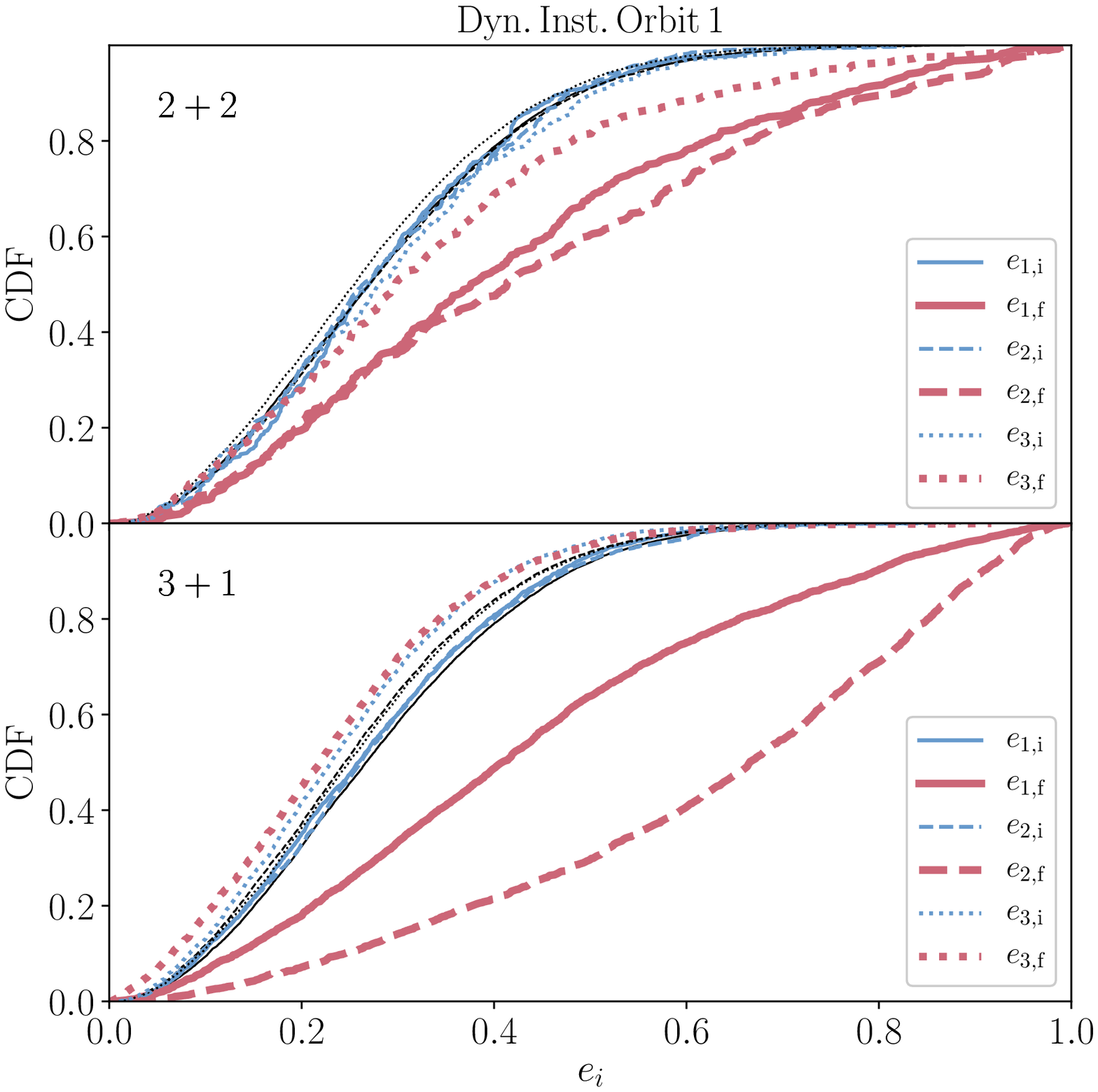}
\includegraphics[scale = 0.45, trim = 10mm -5mm 0mm 10mm]{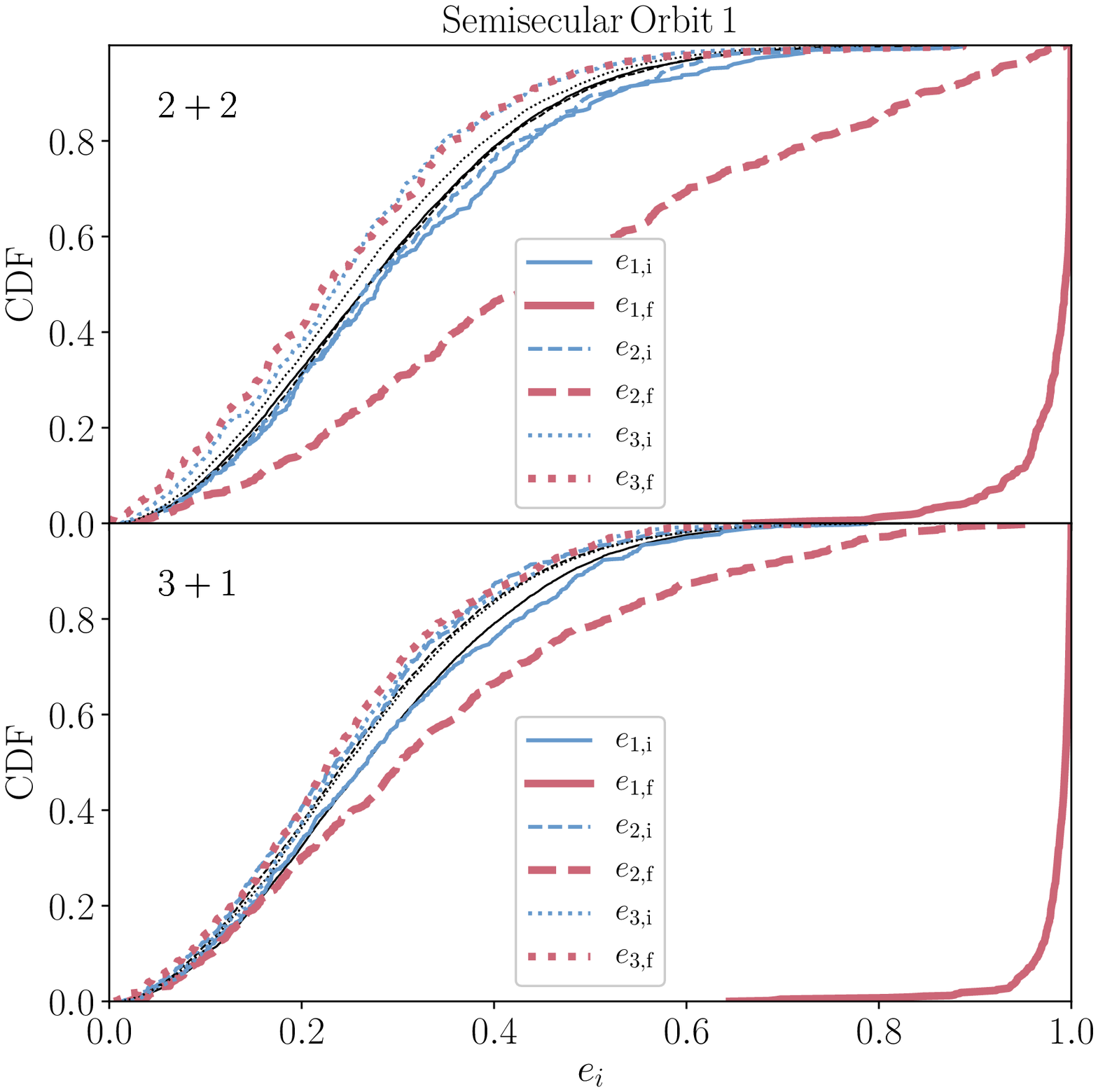}
\caption {Cumulative distributions of the eccentricities for four outcome in the population synthesis simulations, similar to \F\,\ref{fig:pop_syn_sma}. }
\label{fig:pop_syn_e}
\end{figure*}

Here, we focus on the initial properties of the orbits of the systems for the main channels, and on the orbital evolution. In the associated figures, Figures\,\ref{fig:pop_syn_sma}, \ref{fig:pop_syn_e} and \ref{fig:pop_syn_incl}, we plot, for various outcomes, distributions of the semimajor axes, eccentricities and inclinations for the orbits (solid: orbit 1; dashed: orbit 2; dotted: orbit 3), making a distinction between the initial values (blue lines), and values at the end of the simulation (red lines). The thin black lines show the initial distributions for {\it all} systems (i.e., not making a distinction of the outcome), and help to compare the distributions for the different channels across the various figures. In these figures, flybys were taken into account; there generally are only minor differences in the distributions with and without flybys, except for the eccentricity distribution of $e_3$.

\subsubsection{Semimajor axes}
\label{sect:results:orb:sma}
The distributions of the semimajor axes are shown in \F\,\ref{fig:pop_syn_sma} for the no-interaction, RLOF star 1, dynamical instability (orbit 1), and semisecular (orbit 1) outcomes.

As can be expected, the non-interacting systems tend to be those with wide parent orbits (orbit 3 for the 2+2 configuration, and orbits 2 and 3 for the 3+1 configuration) compared to all systems, generally implying weaker secular interactions (i.e., longer LK time-scales). The orbits expand due to mass loss, and this is reflected in the final distributions of all three semimajor axes.

The systems developing RLOF for star 1 tend to have more compact orbits, in particular orbit 1. A more compact orbit 1 implies that it is more likely to trigger RLOF of star 1; to ensure a relatively short secular time-scale, the parent orbit should also be relatively compact. In $\approx 0.4$ of the RLOF star 1 cases, the final $a_1$ is smaller than the initial cutoff (12 $\au$), implying that secular evolution combined with tidal friction has shrunk the orbit before triggering RLOF. Examples of this scenario were given in \S\,\ref{sect:examples:2} and \ref{sect:examples:3}. 

The dynamical instability and semisecular systems (orbit 1) also tend to be more compact compared to all systems. In particular, for these channels and the 2+2 configuration, the inner orbits are significantly wider compared to all systems. For the dynamical instability channel, the inner orbit expands more compared to the semisecular regime channel, indicating that mass loss is the main driver for dynamical instability, whereas secular evolution (i.e., high eccentricities) is the main driver for triggering the semisecular regime.

\subsubsection{Eccentricities}
\label{sect:results:orb:e}

Distributions of the eccentricities are shown for the no-interaction, RLOF star 1, dynamical instability (orbit 1) and semisecular (orbit 1) outcomes in \F\,\ref{fig:pop_syn_e}.

For the non-interacting systems, there is no clear preference for the initial eccentricity of orbits 1 and 2. For 2+2 systems, orbits 1 and 2 become eccentric after 10 Gyr (with very similar distributions), which can be attributed to secular evolution. For 3+1 systems, only orbit 1 becomes eccentric after 10 Gyr, but not orbit 2. This can be understood by noting that a dynamical instability is likely triggered if orbit 2 becomes highly eccentric. The final eccentricities of orbit 2 for the dynamically unstable systems are indeed relatively high. The eccentricity of orbit 3 for the non-interacting systems tends to decrease over time, which can be attributed to the effects of flybys.

The RLOF star 1 systems show a population of circular orbits for orbit 1 at the time of RLOF. This can be understood in conjunction with the population of shrinking orbits in the corresponding distribution of semimajor axes: in these cases, tidal friction shrinks and circularises orbit 1 before triggering RLOF. There is also a significant population ($\approx 0.4$ for the 2+2 configuration, and $\approx 0.2$ for the 3+1 configuration) in which RLOF is triggered in highly eccentric orbits ($e_{1,\mathrm{f}} \gtrsim 0.9$), showing that RLOF in eccentric orbits is important in quadruple systems. The excitation of the eccentricity of orbit 2 at the moment of RLOF can be attributed to secular evolution. 

The semisecular regime is associated with very high eccentricities of orbit 1 (median $\gtrsim 0.99$), driven by secular evolution. There is no strong dependence of the initial eccentricity for this channel.

\begin{figure*}
\center
\includegraphics[scale = 0.45, trim = 10mm 5mm 0mm 10mm]{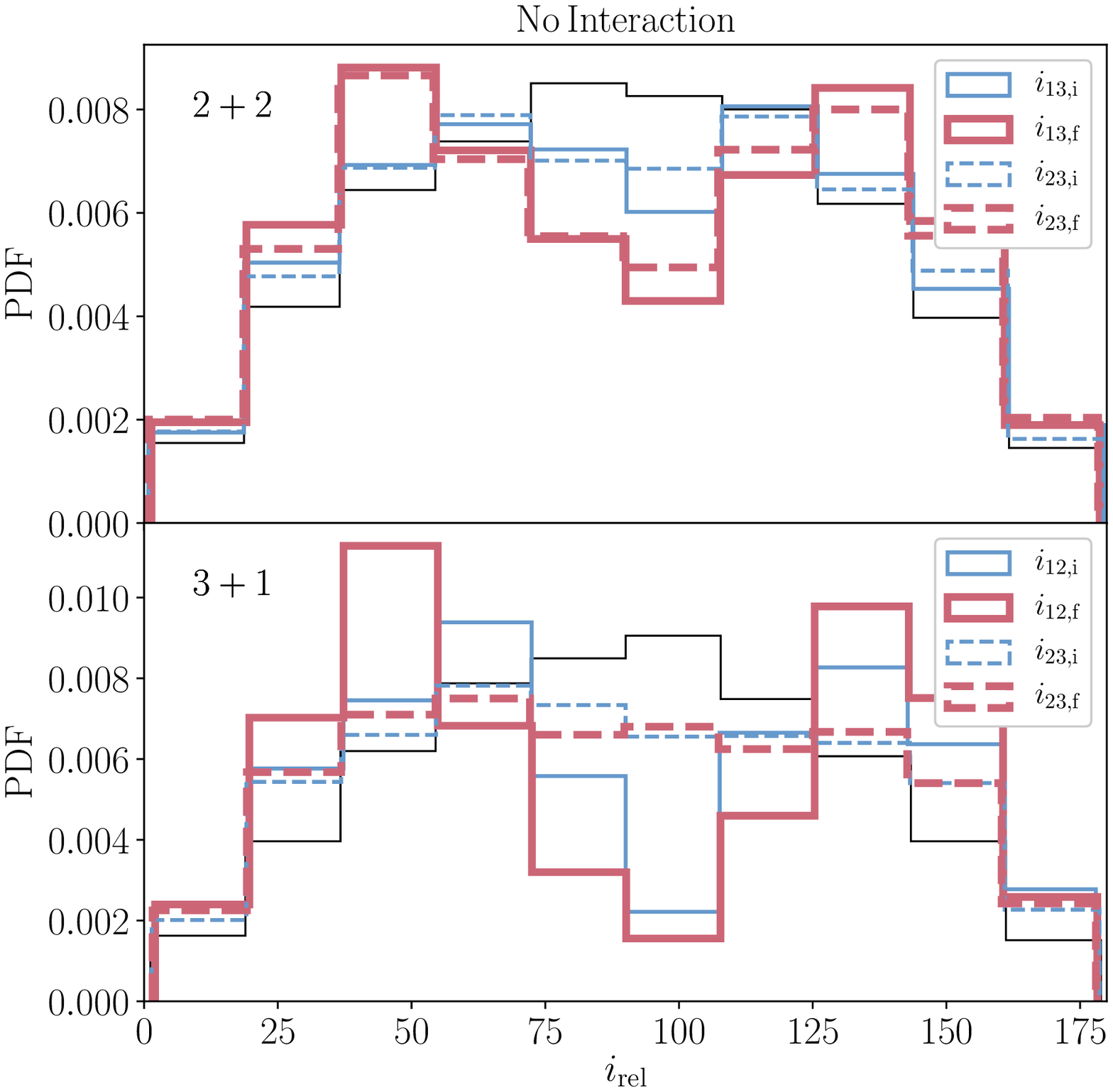}
\includegraphics[scale = 0.45, trim = 10mm 5mm 0mm 10mm]{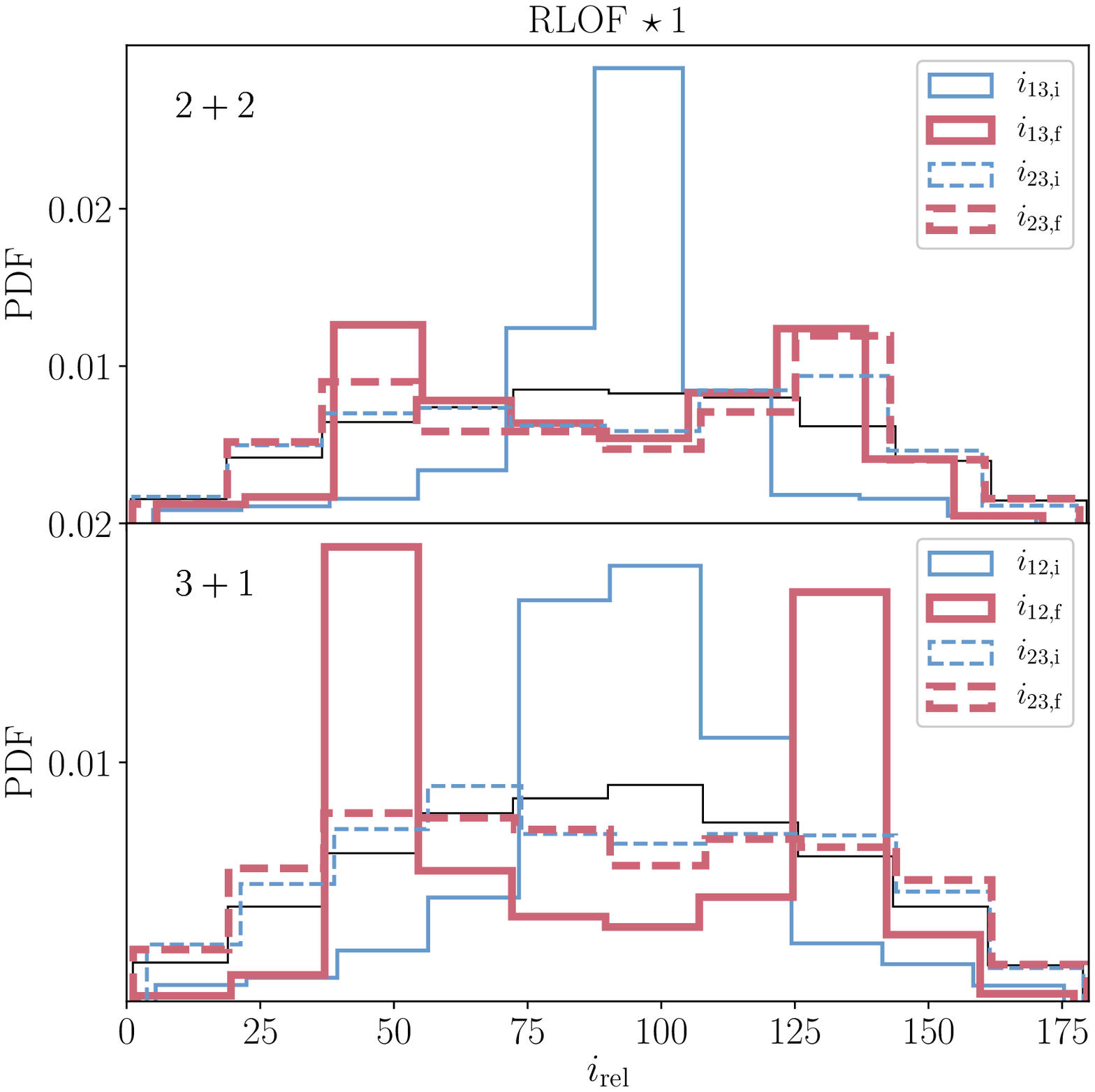}
\includegraphics[scale = 0.45, trim = 10mm 5mm 0mm 10mm]{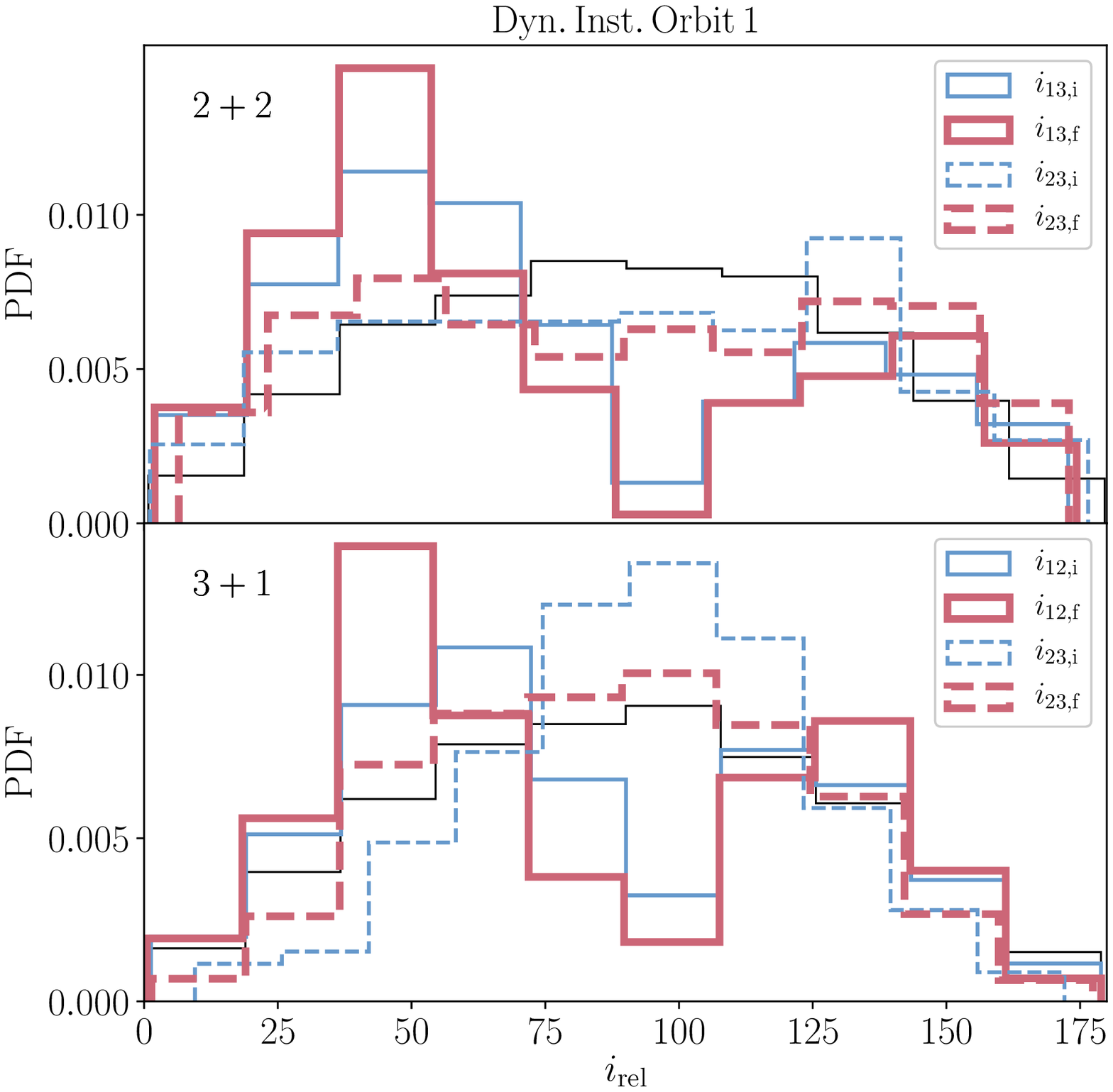}
\includegraphics[scale = 0.45, trim = 10mm 5mm 0mm 10mm]{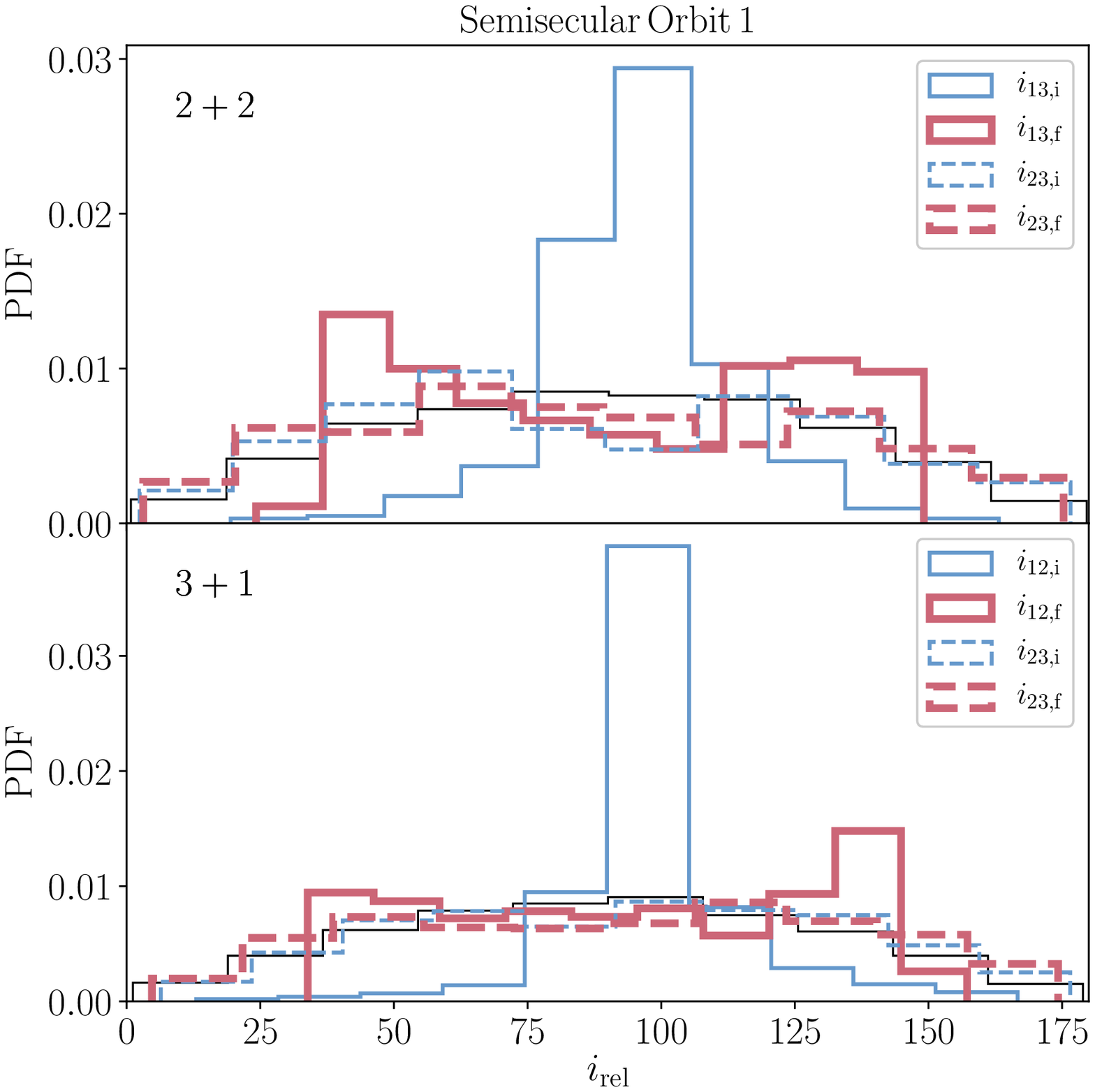}
\caption {Probability density distributions of the inclinations relative to the parent for four outcome in the population synthesis simulations, similar to \F\,\ref{fig:pop_syn_sma}. }
\label{fig:pop_syn_incl}
\end{figure*}

\subsubsection{Inclinations}
Distributions of the inclinations relative to the parents of orbits 1 and 2 are shown for the no-interaction, RLOF star 1, dynamical instability (orbit 1) and semisecular (orbit 1) outcomes in \F\,\ref{fig:pop_syn_incl}. Note that for the 2+2 configuration, these inclinations are $i_{13}$ and $i_{23}$ for orbits 1 and 2, respectively; for the 3+1 configuration, they are $i_{12}$ and $i_{23}$. We recall that the initial distributions of these mutual inclinations were assumed to be flat in their cosine ($\mathrm{d}N/di_{ij} \propto \sin i_{ij}$), i.e., corresponding to random mutual orientations for all orbits. 

The non-interacting systems show a paucity of inclinations near $90^\circ$. This can be understood by noting that such high inclinations tend to trigger interactions through secular interactions, which is strongly reflected in the distributions of the initial inclinations of orbit 1 relative to its parent for the RLOF star 1 and semisecular outcomes. The final distributions of the latter inclinations are peaked around $\approx 40^\circ$ and $\approx 130^\circ$, which is the characteristic LK angle at which the eccentricity oscillations are at maximum amplitude. The semisecular channel shows a particularly strong preference for high initial inclinations.

\begin{figure}
\center
\includegraphics[scale = 0.45, trim = 10mm 5mm 0mm 10mm]{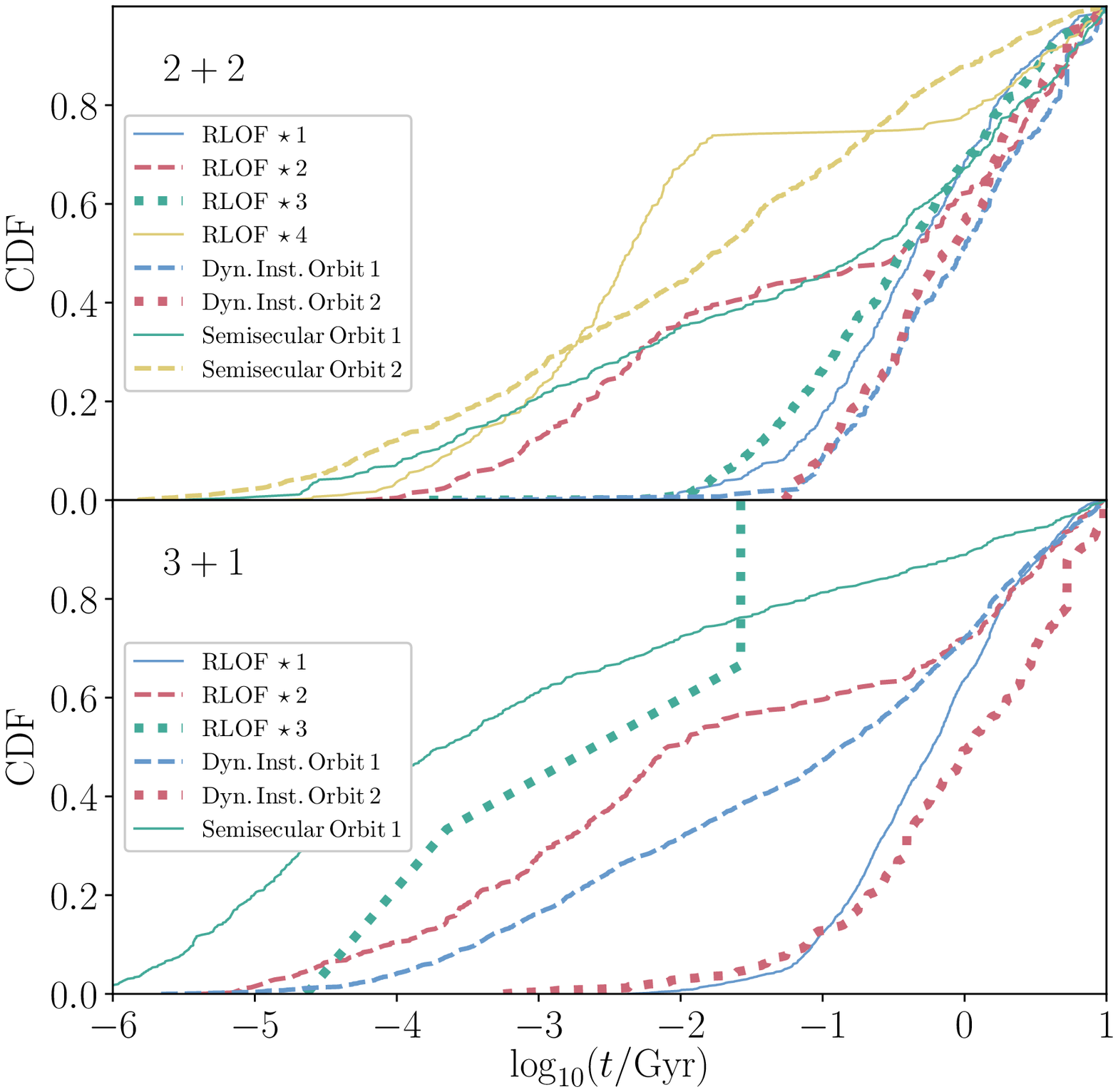}
\caption {Cumulative distributions of the stopping times of several channels in the simulations for the 2+2 (top panel) and 3+1 (bottom panel) configurations. Refer to the legend for the meaning of the colours and line styles. }
\label{fig:stop_times}
\end{figure}

\subsection{Stopping times}
\label{sect:results:time}
In \F\,\ref{fig:stop_times}, we show the cumulative distributions of the stopping times of several channels in the simulations for the 2+2 (top panel) and 3+1 (bottom panel) configurations. 

RLOF in the simulations tends to occurs relatively early, typically before 1 Gyr. The semisecular regime can be triggered very early, although delay times approaching 10 Gyr are also possible. Note that in the early semisecular channel the stars are still MS stars, and are therefore of no interest for SNe Ia. Dynamical instability in orbit 1 can occur at a wide range of ages, from only $10^4\,\mathrm{yr}$ (for the 3+1 configuration), to several Gyr. In the case of 3+1 systems, early dynamical instability of orbit 1 is driven by increased eccentricity of orbit 2 due to LK oscillations with orbit 3. Later instability can again be driven by enhanced $e_3$, but also due to mass-loss-driven orbital expansion.

\subsection{SNe Ia from WD collisions}
\label{sect:results:col}

\subsubsection{Channels}
\label{sect:results:col:chan}
As mentioned in \S\,\ref{sect:introduction}, colliding CO WDs are potentially efficient progenitors of SNe Ia (e.g., \citealt{2009MNRAS.399L.156R,2009ApJ...705L.128R,2010ApJ...724..111R,2012ApJ...747L..10P,2015ApJ...807..105S}). We consider the following channels found in our simulations for collisions of CO WDs.

\begin{enumerate}
\item Direct collisions of two CO WDs found during the secular integrations (in orbits 1 or 2 for the 2+2 systems, or orbit 1 for the 3+1 systems). These can be considered to be `secular' collisions, in the sense that semisecular evolution was not important for driving the high eccentricity needed for collision (otherwise, the simulation would have stopped and been flagged as a semisecular system prior to collision). Note that we also included tidal evolution for WDs in the integrations (i.e., degenerate damping, as part of the prescription of \citealt{2002MNRAS.329..897H}), and energy and angular-momentum loss due to gravitational wave radiation.
\item Collisions of two CO WDs resulting from a dynamical instability. We consider the cases when dynamical instability occurs with two CO WDs in either orbit 1 or 2 (2+2 systems), or orbit 1 only (3+1 systems). We expect collisions to occur after dynamical instability in a fraction of cases. In the $N$-body simulations of \citet{2012ApJ...760...99P}, it was found for triples that the collision probability is about 10\%. Here, for simplicity, we adopt the same efficiency, with the caveat that the collision efficiency could conceivably be higher in quadruple systems compared to triple systems. In any case, our rates from this channel can easily be adjusted to take into account a higher collision efficiency. Moreover, even if the efficiency is 100\%, then the SNe Ia rates would still be about a factor 100 times lower than the observed rates (see below). 
\item Collisions of two CO WDs after entering the semisecular regime (orbits 1 and 2 for the 2+2 configuration, and orbit 1 for the 3+1 configuration). When a system enters the semisecular regime, a collision is expected to occur after a certain delay time \citep{2012arXiv1211.4584K}. We adopt a mean delay time of \citep{2012arXiv1211.4584K}
\begin{align}
\label{eq:t_col}
\overline{t}_{\mathrm{col},i} = \frac{a_i}{4 R_\mathrm{WD}} P_{\mathrm{orb},i},
\end{align}
where $i$ refers to either orbit 1 or 2, and $R_\mathrm{WD}$ is the radius of star 1 (orbit 1) or star 3 (orbit 2). For a given CO WD semisecular system, we assume that a CO WD collision occurs (with 100\% efficiency) at a time given by the time of entering the semisecular regime, plus $\overline{t}_{\mathrm{col},i}$. Below, for illustrative purposes, we also give the rates without taking into account equation~(\ref{eq:t_col}), which we refer to as the `uncorrected' rates.
\end{enumerate}

\subsubsection{Normalisation}
\label{sect:results:col:norm}
We normalise our rates as follows. Our procedure, which is similar to that of \citet{2013MNRAS.430.2262H}, is to compute the total mass $M_\mathrm{tot}$ of the population represented by the $N_\mathrm{calc}=10^4$ calculated systems. Let the galactic (field) population consist of $N_\mathrm{tot}$ gravitationally bound systems with $N_\mathrm{bin} = \alpha_\mathrm{bin} N_\mathrm{tot}$ binaries, $N_\mathrm{tr} = \alpha_\mathrm{tr} N_\mathrm{tot}$ triples, $N_\mathrm{2+2} = \alpha_\mathrm{2+2} N_\mathrm{tot}$ quadruples in the 2+2 configuration, $N_\mathrm{3+1} = \alpha_\mathrm{3+1} N_\mathrm{tot}$ quadruples in the 3+1 configuration, and $(1-\alpha_\mathrm{bin} - \alpha_\mathrm{tr} - \alpha_\mathrm{2+2} - \alpha_\mathrm{3+1}) N_\mathrm{tot}$ single stars (we ignore quintuple and higher-order systems). We assume a binary fraction of $\alpha_\mathrm{bin}=0.6$, a triple fraction of $\alpha_\mathrm{tr} = 0.25$, a 2+2 quadruple fraction of $\alpha_{2+2} = (2/3) \times 0.03$, and a 2+2 quadruple fraction of $\alpha_{3+1} = (1/3) \times 0.03$. The binary, triple and total quadruple fractions are (loosely) adopted from \citet{2010ApJS..190....1R}; the ratio of the 2+2 to 3+1 systems is taken to be 2:1, consistent with F and G dwarfs in the Solar neighbourhood \citep{2014AJ....147...86T,2014AJ....147...87T}. 

The number of calculated quadruple systems constitutes a fraction $f_\mathrm{calc,2+2}$ ($f_\mathrm{calc,3+1}$) of all $N_\mathrm{2+2}$ ($N_\mathrm{3+1}$) quadruple systems in the 2+2 (3+1) configuration. We compute these calculated fractions for each configuration by sampling systems satisfying stability criteria and using the assumptions of \S\,\ref{sect:IC:orbits}, and determining the number of the subset of those systems that satisfy the constraints that were used in the population synthesis calculations. The latter constraints are: the primary mass $1<m_1/\msun<6.5$ (as opposed to $0.1<m_1/\msun<80$ for all systems), the secondary mass $m_2 > 1\,\msun$ (as opposed to $m_2>0.1\,\msun$ for all systems), and the semilatus rectum $a_i > 12 \, \au$ ($i\in \{1,2\}$ for 2+2, and $i=1$ for 3+1 systems, as opposed to $0<\mathrm{log}_{10}[P_{\mathrm{orb},i}/\mathrm{d}]<10$ for all systems). We then find $f_\mathrm{calc,2+2} \simeq 0.030$ and $f_\mathrm{calc,3+1} \simeq 0.015$.

The total mass of all single, binary, triple and quadruple systems is then estimated as follows. First, note that the average mass assuming a Kroupa mass distribution is $M_\mathrm{Kr} \approx 0.5006\,\msun$ (see, e.g., equation A4 of \citealt{2013MNRAS.430.2262H}). Assuming flat mass ratio distributions as in \S\,\ref{sect:IC:masses}, we then estimate the mass of all single stars to be $M_\mathrm{s} \approx M_\mathrm{Kr} N_\mathrm{s}$, of all binary stars $M_\mathrm{bin} \approx \left(1+\frac{1}{2} \right)M_\mathrm{Kr} N_\mathrm{bin} = \frac{3}{2} M_\mathrm{Kr} N_\mathrm{bin}$, of all triple stars $M_\mathrm{tr} \approx \left(\frac{3}{2}+\frac{1}{2} \frac{3}{2}\right)M_\mathrm{Kr} N_\mathrm{tr} = \frac{9}{4} M_\mathrm{Kr} N_\mathrm{tr}$, for all 2+2 quadruple stars $M_\mathrm{2+2} \approx \left(\frac{3}{2}+\frac{1}{2} \frac{3}{2}\right)M_\mathrm{Kr} N_\mathrm{2+2} = \frac{9}{4} M_\mathrm{Kr} N_\mathrm{2+2}$, and for all 3+1 quadruple stars $M_\mathrm{3+1} \approx \left(\frac{3}{2}+\frac{1}{2} \frac{3}{2} + \frac{1}{2} \frac{1}{2} \frac{3}{2} \right)M_\mathrm{Kr} N_\mathrm{3+1} = \frac{21}{8} M_\mathrm{Kr} N_\mathrm{3+1}$. This gives a total mass
\begin{align}
\nonumber M_{\mathrm{tot},C} &\approx M_\mathrm{Kr} \left (N_\mathrm{s} + N_\mathrm{bin} + N_\mathrm{tr} + N_\mathrm{2+2} + N_\mathrm{3+1} \right ) \\
&= M_\mathrm{Kr} \frac{N_\mathrm{calc}}{f_{\mathrm{calc},C}} \left (1 + \frac{3}{2} \alpha_\mathrm{bin} + \frac{5}{4} \alpha_\mathrm{tr} + \frac{5}{4} \alpha_{2+2} + \frac{13}{8} \alpha_{3+1} \right ),
\end{align}
where $C$ refers to either the 2+2 or 3+1 configurations. Substituting the above numbers, we find $M_\mathrm{tot,2+2} \approx 1.4 \times 10^7\,\msun$, and  $M_\mathrm{tot,3+1} \approx 5.4 \times 10^7\,\msun$. These masses are used to normalise the rates presented below in \S\,\ref{sect:results:col:DTD}.

\begin{figure}
\center
\includegraphics[scale = 0.45, trim = 0mm 10mm 0mm 25mm]{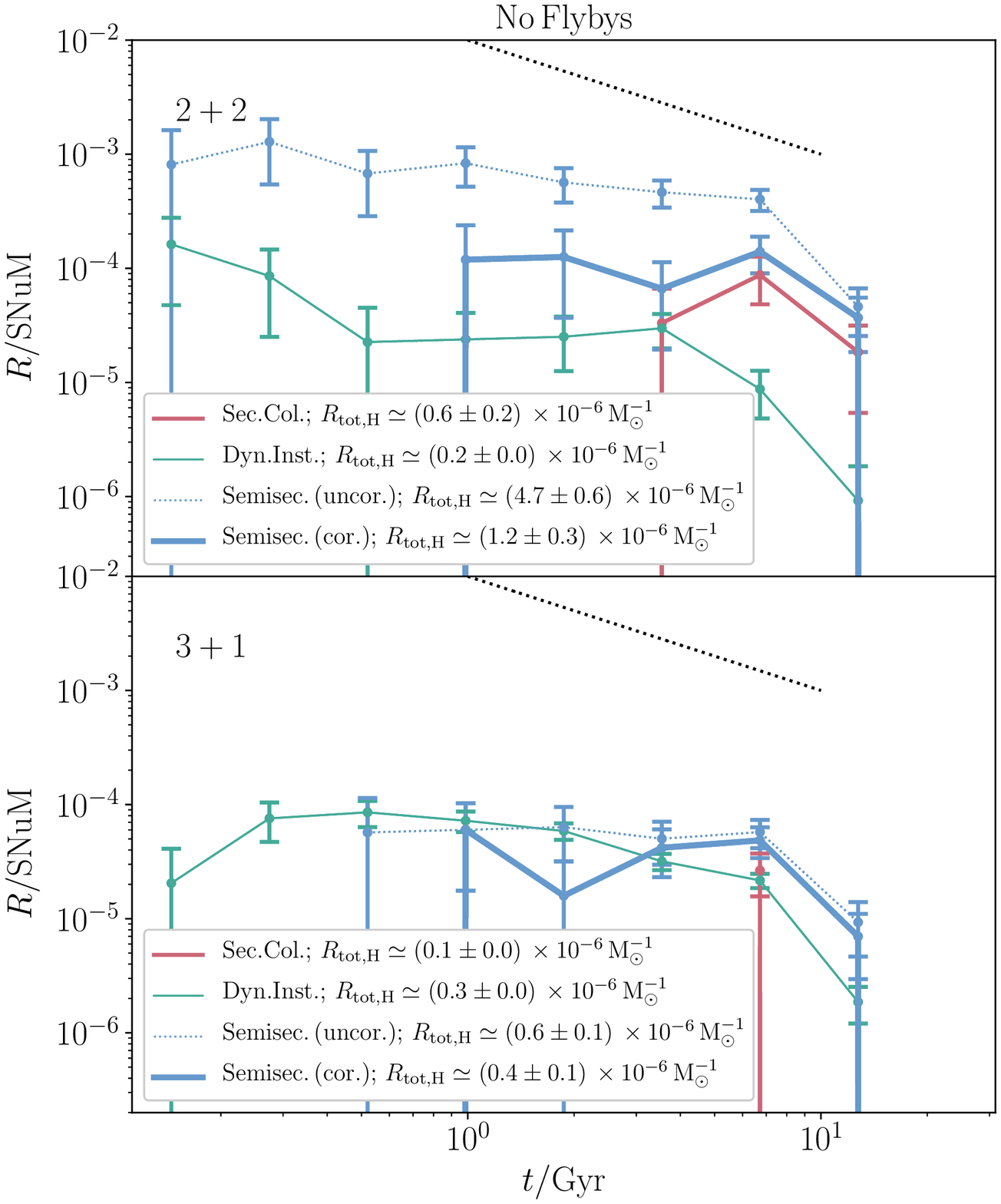}
\includegraphics[scale = 0.45, trim = 0mm 20mm 0mm 0mm]{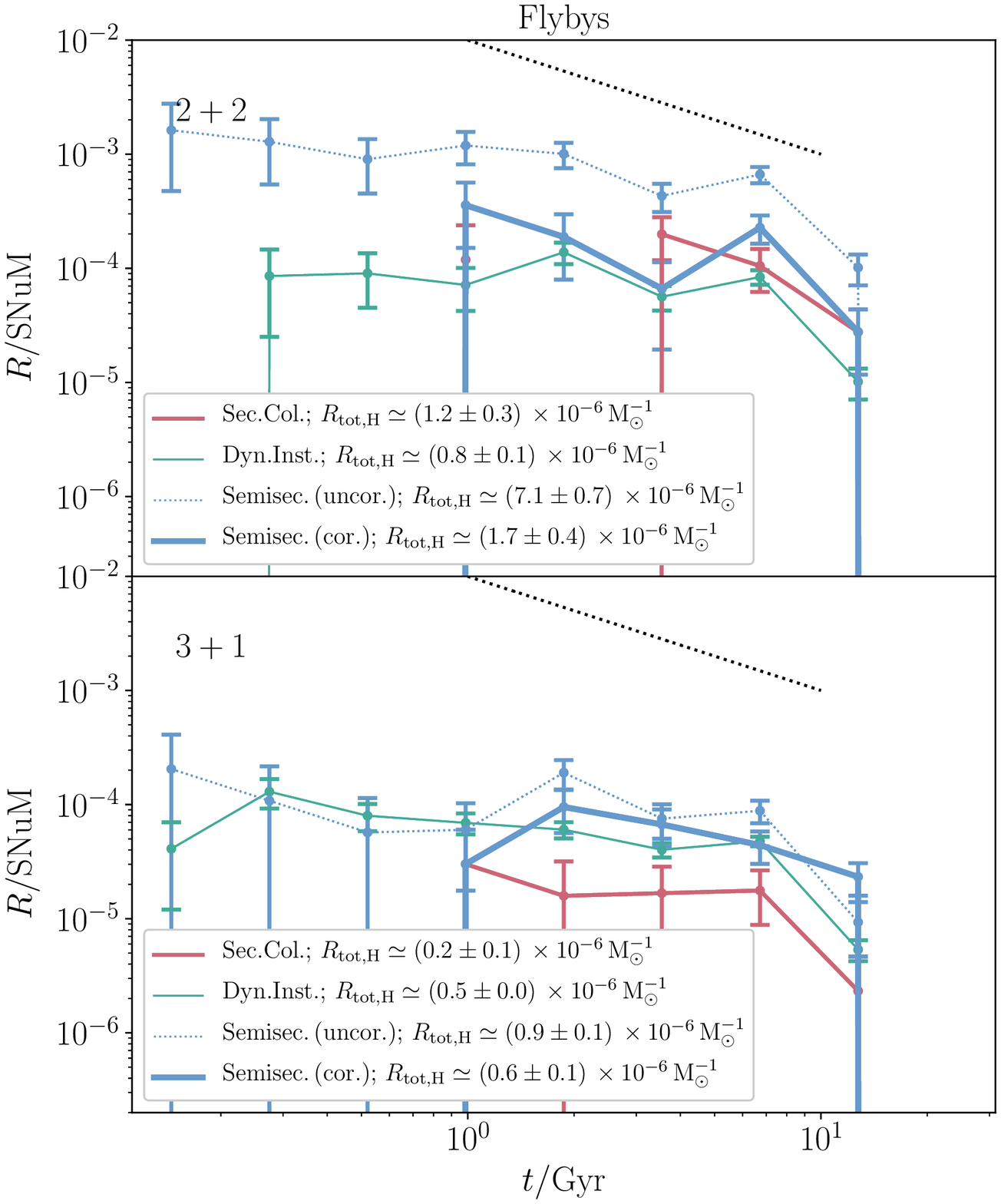}
\caption { DTDs for collision-induced SNe Ia. The DTDs are expressed in units of SNuM, i.e., rate per century per $10^{10}\,\msun$ of Solar mass. The top two panels apply to the 2+2 and 3+1 configurations without the inclusion of flybys, whereas flybys are included in the bottom two panels. We make a distinction between the following channels: secular collisions (solid red lines), dynamical instability (solid green lines), and semisecular regime (uncorrected: blue dotted lines, and corrected: solid blue lines). The black dotted line shows a dependence $R\propto t^{-1}$. The time-integrated rates (with times less than $t_\mathrm{H} \equiv 14\,\mathrm{Gyr}$) are given in the labels. }
\label{fig:DTD}
\end{figure}

\begin{figure}
\center
\includegraphics[scale = 0.45, trim = 0mm 0mm 0mm 10mm]{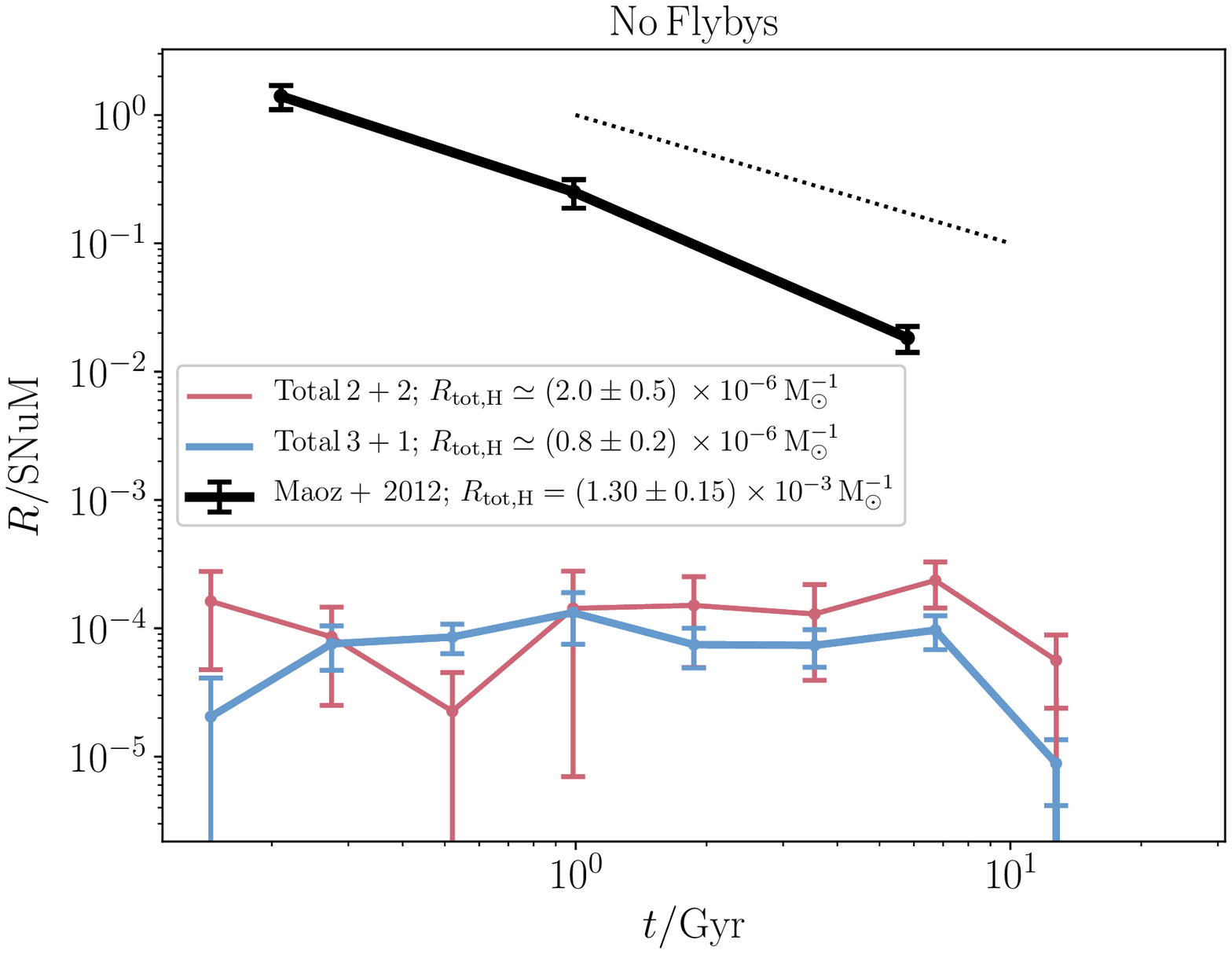}
\includegraphics[scale = 0.45, trim = 0mm 0mm 0mm 0mm]{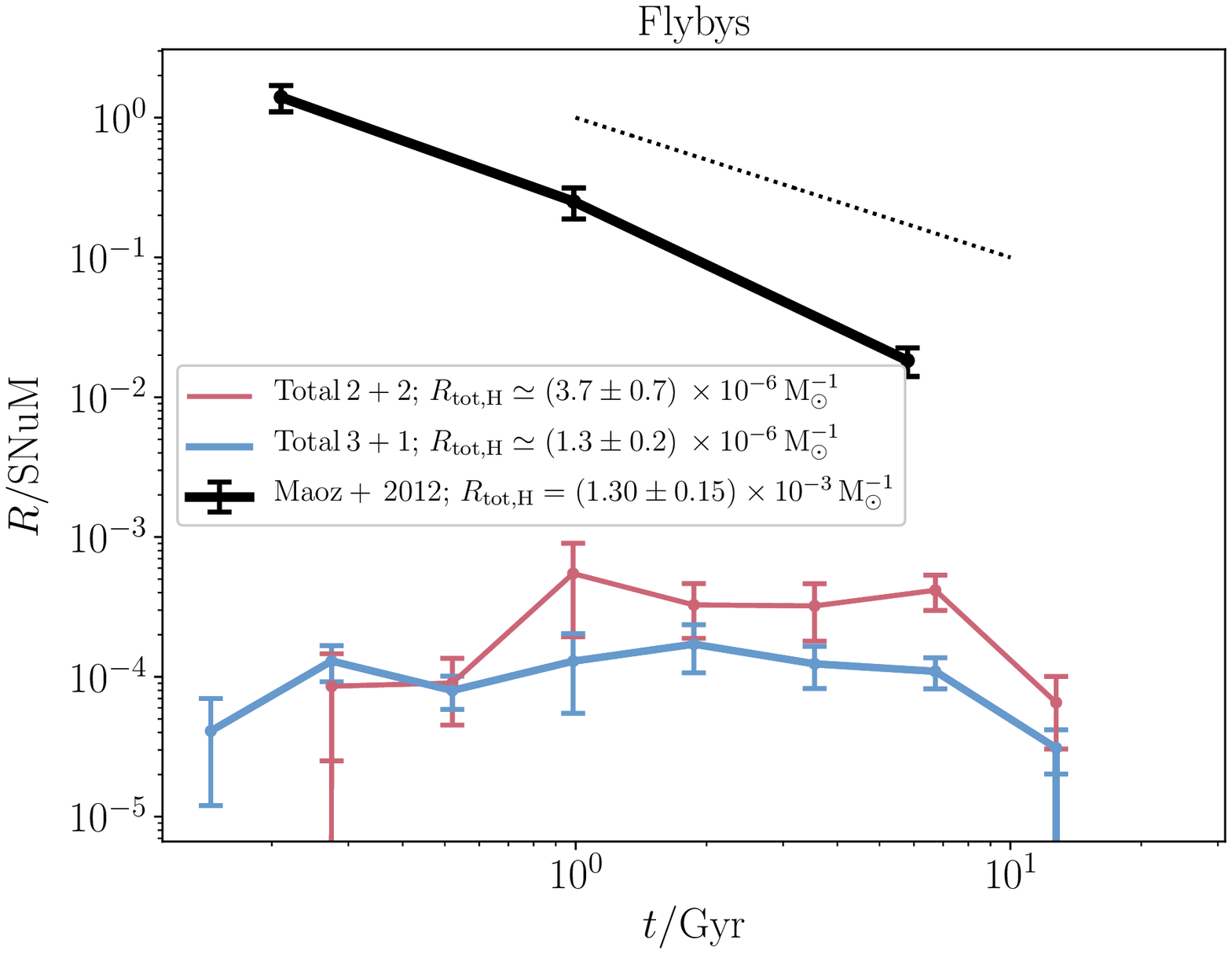}
\caption {DTDs for the combined channels and separately for the 2+2 (red line) and 3+1 (blue line) systems, with flybys excluded (top panel), and included (bottom panel).  The black solid line shows observations from \citet{2012MNRAS.426.3282M}, and the black dotted line shows a dependence $R\propto t^{-1}$. The time-integrated rates are given in the labels. }
\label{fig:DTD_all}
\end{figure}

\subsubsection{Delay-time distribution}
\label{sect:results:col:DTD}
In \F\,\ref{fig:DTD}, we show the delay-time distributions (DTDs) for the channels described in \S\,\ref{sect:results:col:chan} and using the normalisation discussed in \S\,\ref{sect:results:col:norm}. Here, we assume that all CO WD collisions lead to SNe Ia (note that we assume that 10\% of the CO WD dynamical instability systems lead to collisions). This is an optimistic assumption, given that not all collisions may result in an SNe Ia explosion (e.g., \citealt{2009MNRAS.399L.156R,2009ApJ...705L.128R,2010ApJ...724..111R,2012ApJ...747L..10P,2015ApJ...807..105S}). The DTDs are expressed in units of SNuM, i.e., rate per century per $10^{10}\,\msun$ of Solar mass. The top two panels apply to the 2+2 and 3+1 configurations without the inclusion of flybys, whereas flybys are included in the bottom two panels. 

The DTDs for the various channels are fairly flat: flatter than $R\propto t^{-1}$, which is often found in binary population synthesis studies (e.g., \citealt{2012A&A...546A..70T}), and observations (e.g., \citealt{2012MNRAS.426.3282M}). We note that flat DTDs are also found for collision-induced CO WDs in triple systems \citep{2017arXiv170900422T}. 

The time-integrated rates are given in the labels in \F\,\ref{fig:DTD}, where the error is based on Poisson statistics, and we only consider delay times less than $t_\mathrm{H} \equiv 14\,\mathrm{Gyr}$ (the latter is important for the `corrected' semisecular systems, some of which give a collision time later than $t_\mathrm{H}$). Typically, time-integrated rates are on the order of $10^{-6}\,\mathrm{M}_\odot^{-1}$, similar to the triple rates \citep{2013MNRAS.430.2262H,2017arXiv170900422T}. It is important to take into account the delay-time correction for the semisecular systems; the difference in the time-integrated rate between the corrected and uncorrected rates for the 2+2 systems is a factor of $\approx 4$. The effects of flybys are relatively important: the inclusion of flybys can increase the rates by factors of up to $\approx 2$. 

In \F\,\ref{fig:DTD_all}, we add the contributions from all channels and show the DTDs for the 2+2 and 3+1 configurations. We also include three observational data points from \citet{2012MNRAS.426.3282M}. Clearly, the observed rates are much higher than those found in our simulations: the observed integrated rate is of order $10^{-3}\,\mathrm{M}_\odot^{-1}$, whereas the simulated quadruple integrated rate is of order $10^{-6}\,\mathrm{M}_\odot^{-1}$. In addition, the simulated DTDs are significantly more flat compared to the observed DTD.

\section{Discussion}
\label{sect:discussion}

\subsection{Tighter systems and RLOF}
\label{sect:discussion:RLOF}
We considered a subset of all quadruple systems by restricting to systems with semilatus recti larger than $12\,\au$ (i.e., not interacting in binary isolation). There may be a contribution of collision-induced SNe Ia from tighter systems, and we did not consider here. 

In addition, we stopped our simulations at the onset of RLOF (see \S\,\ref{sect:meth:sc}). Typically, RLOF is expected to lead to CE evolution and significant shrinkage of the orbit, such that further secular evolution is suppressed and a collision of CO WDs in an eccentric orbit is avoided, although a circular merger after double WD formation is still possible \citep{2013MNRAS.430.2262H}. However, RLOF can also occur in eccentric systems as found in our simulations (see \S\,\ref{sect:results:orb:e}), and it is not clear what the result would be in this case. RLOF in eccentric systems that are also driven by secular evolution is beyond the scope of this work, but certainly merits further investigation. 

With the above in mind, we emphasise that our rates, although very low compared to observations, should be interpreted as lower limits.

\subsection{Suborbital effects}
\label{sect:discussion:sub}
In our integrations, we neglected suborbital effects by averaging the equations of motion over all three orbits. This approximation can break down in cases when the time-scale for the angular momentum or eccentricity vector to change is comparable to some of the orbital periods. For example, when an inner orbit precession frequency is comparable to the outer orbital period, the evection resonance may come into play and modulate the secular eccentricity oscillations \citep{2005MNRAS.358.1361I,2014MNRAS.439.1079A,2016MNRAS.458.3060L,2017arXiv170908682F}. Another example is the occurrence of mean-motion resonance in 2+2 systems \citep{2018MNRAS.475.5215B}. 

It is left for future work to quantify the importance of these suborbital effects for the collision-induced SNe Ia rates in quadruples. However, in our view it is hard to imagine that these effects could increase the rates by three orders of magnitude, which is necessary to approach the observed rates.

\subsection{Galactic tides}
\label{sect:discussion:gal_tide}
We did not consider the effects of galactic tides in our simulations. Galactic tides are typically less important compared to flybys at $\sim 10^4\,\au$, the typical initial semimajor axis of orbit 3 (see \S\,\ref{sect:IC}). However, they are typically more important for orbits approaching $10^5\,\au$, and some of the outermost orbits in our simulations do evolve to $\sim 10^5\,\au$ due to mass-loss-induced orbital expansion. It is left for future work to evaluate the importance of galactic tides on SNe Ia rates in quadruples.

\section{Conclusions}
\label{sect:conclusions}
We studied the evolution of hierarchical quadruple star systems in the 2+2 (two binaries orbiting each other's barycentre) and 3+1 (triple orbited by a fourth star) configurations. We took into account the effects of secular dynamical evolution, stellar evolution, tidal evolution and encounters with passing stars, and we focused on SNe Ia driven by collisions of CO WDs. Our main conclusions are given below.

\medskip \noindent 1. The quadruple systems considered here are initially wide (initial semilatus recti larger than $12\,\au$), implying no interaction if the orbits were isolated. However, taking into account the dynamical evolution, we found that $\approx 0.4$ ($\approx 0.6$) of 2+2 (3+1) systems interact during their evolution. In particular, RLOF, possibly in highly eccentric orbits, is triggered in a fraction of about 0.2 (0.4) of cases for 2+2 (3+1) systems. In addition, dynamical instability can be triggered due to mass-loss-driven orbital expansion or secular evolution (the latter mainly in 3+1 systems), or a semisecular regime can be entered in which the orbit-averaged equations of motion break down, and a collision could occur after a certain delay time. 

\medskip \noindent 2. In a significant fraction of systems, an interaction occurs before the formation of two CO WDs, thereby avoiding WD collisions in later stages. For example, the fraction of semisecular systems is $\approx 0.04$ ($\approx 0.06$) for 2+2 (3+1) systems, and in the majority ($\approx 60\%$ for 2+2 systems, and $\approx 90\%$ for 3+1 systems) of these, the semisecular regime is entered when the stars are still MS stars. This shows that it is important to take into account the pre-WD evolution when considering WD collision-induced SNe Ia in quadruple systems.

\medskip \noindent 3. We identified a number of channels that can result in CO WD mergers, and, therefore, potentially SNe Ia. These channels include direct collisions due to extremely high eccentricities according to the averaged equations of motion, collisions following dynamical instability, and collisions due to evolution in the semisecular regime.

\medskip \noindent 4. We computed from our simulations the delay-time distributions (DTDs) of CO WD collision-induced SNe Ia. The DTDs are fairly flat (somewhat flatter than $\propto t^{-1}$). The time-integrated SNe Ia rate combined from all channels in our simulations is $(3.7\pm0.7) \times 10^{-6} \,\mathrm{M}_\odot^{-1}$ and $(1.3\pm0.2) \times 10^{-6} \,\mathrm{M}_\odot^{-1}$ for 2+2 and 3+1 systems, respectively. These rates are about three orders of magnitude lower compared to the observed rate, $(1.30 \pm 0.15) \times 10^{-3} \, \mathrm{M}_\odot^{-1}$ \citep{2012MNRAS.426.3282M}. However, we emphasise that our rates are lower limits given that we considered wide systems only and assumed that RLOF does not lead to collision-induced CO WD mergers. 

\medskip \noindent 5. Although flybys are not important for the outcomes of the main channels (in particular, no interaction vs. interaction), they are important for the collision-induced SNe Ia rates. We found that if flybys are not taken into account, the rates decrease by a factor of up to $\approx 2$.

\section*{Acknowledgements}
I thank Daniel Fabrycky for stimulating discussions, and the anonymous referee for a helpful report. I gratefully acknowledge support from the Institute for Advanced Study and The Peter Svennilson Membership.

\bibliographystyle{mnras}
\bibliography{literature}

\label{lastpage}
\end{document}